\documentclass[12pt, draftclsnofoot, onecolumn]{IEEEtran}
\usepackage{xspace,amsmath,amssymb,amsfonts,epsfig,syntonly}
\usepackage{cite,bm,color,url,textcomp,empheq,boxedminipage}
\usepackage{algorithmicx,algorithm,algpseudocode}
\usepackage{epstopdf}
\usepackage{empheq}

\usepackage{graphicx,graphics}  % Written by David Carlisle and Sebastian Rahtz
\usepackage{multirow,multicol}
\usepackage{psfrag}    % Written by Craig Barratt, Michael C. Grant,
\usepackage{stfloats}

\newtheorem{lemma}{\hspace{-11pt}\bf Lemma}

\newtheorem{theorem}{\hspace{-11pt}\bf Theorem}

\newtheorem{fact}{\hspace{-11pt}\bf Fact}
\newtheorem{remark}{\hspace{-11pt}\bf Remark}

\long\def\symbolfootnote[#1]#2{\begingroup
\def\thefootnote{\fnsymbol{footnote}}
\footnote[#1]{#2}\endgroup}

\psfull

\allowdisplaybreaks[4]

\begin{document}

%%%%%%%%%%%%%%%%%%%%%%%%%%%%%%%%%%%%%%%%%%%%%%%%%%%%%%%%%%%%%%%%%%%%%%
%                                                                    %
%               Begin Document                                       %
%                                                                    %
%%%%%%%%%%%%%%%%%%%%%%%%%%%%%%%%%%%%%%%%%%%%%%%%%%%%%%%%%%%%%%%%%%%%%%

%\title{General Rank-$K$ Beamforming Design for Multiuser MISO Downlinks}
%\title{ SDP Relaxation Performance Achievable Transmit Beamforming Design for MU-MISO Downlinks}
%\title{MU-MISO Downlink Beamforming Optimization with Transmit Covariance Rotation}
%\title{Structured But Optimal Beamforming Design for MU-MISO Downlinks}
\title{REEL-BF Design: Achieving the SDP Bound for Downlink Beamforming with Arbitrary Shaping Constraints}

\author{Feng Wang, Chongbin Xu, Yongwei Huang, Xin Wang, and Xiqi Gao

\thanks{Part of this paper has been presented in the IEEE Global Communications Conference (GLOBECOM), Washington, D.C., USA, December 4--8, 2016.}

%\thanks{This work was supported in part by China Recruitment Program of Global Young Experts, the Program for New Century Excellent Talents in University, and the Innovation Program of Shanghai Municipal Education Commission.}
%

\thanks{F. Wang, C. Xu, and X. Wang are with the Key Laboratory for Information Science of Electromagnetic Waves (MoE), the Department of Communication Science and Engineering, Fudan University, Shanghai, China (e-mail: fengwang.nl@gmail.com, chbinxu@fudan.edu.cn, xwang11@fudan.edu.cn). Xin Wang is the corresponding author.}
\thanks{Y. Huang is with the Department of Mathematics, Hong Kong Baptist University, Kowloon, Hong Kong (e-mail: huang@math.hkbu.edu.hk).}
\thanks{X. Q. Gao is with the National Mobile Communications Research Laboratory, Southeast University, Nanjing, China (e-mail: xqgao@seu.edu.cn).}

}

\maketitle

\begin{abstract}
This paper considers the beamforming design for a multiuser multiple-input single-output (MISO) downlink with an arbitrary number of (context-specific) shaping constraints. In this setup, the state-of-the-art beamforming schemes cannot attain the well-known performance bound promised by the semidefinite program (SDP) relaxation technique. To close the gap, we propose a redundant-signal embedded linear beamforming (REEL-BF) scheme, where each user is assigned with one information beamformer and several shaping beamformers. It is shown that the proposed REEL-BF scheme can perform general rank-$K$ beamforming for user symbols in a low-complexity and structured manner. In addition, sufficient conditions are derived to guarantee that the REEL-BF scheme always achieves the SDP bound for linear beamforming schemes. Based on such conditions, an efficient algorithm is then developed to obtain the optimal REEL-BF solution in polynomial time. Numerical results demonstrate that the proposed scheme enjoys substantial performance gains over the existing alternatives.
\end{abstract}

\begin{IEEEkeywords}
Downlink beamforming, shaping constraints, unitary rotation, semidefinite program relaxation.
\end{IEEEkeywords}

%%%%%%%%%%%%%%%%%%%%%%%%%%%%%%%%%%%%%%%%%%%%%%%%%%%%%%%%%%%%%%%%%%%%%%
%                                                                    %
%               Section: Introduction                                %
%                                                                    %
%%%%%%%%%%%%%%%%%%%%%%%%%%%%%%%%%%%%%%%%%%%%%%%%%%%%%%%%%%%%%%%%%%%%%%

\section{Introduction}

Linear transmit-beamforming is a low-complexity strategy to reduce the co-channel interference and increase energy/spectrum efficiency for multiuser multiple-input single-output (MU-MISO) downlinks \cite{Ben01, Bjo14, Sch04,Ger10}. A typical beamforming design is to minimize the transmission power subject to the received signal-to-interference-plus-noise ratio (SINR) constraints per user. It was shown that this beamforming problem can be relaxed to a semidefinite program (SDP) that can be efficiently solved in polynomial time \cite{Luo10}. In addition, it was established in \cite{Ben01,Bjo14, Sch04,Ger10,Luo10} that such SDP relaxations always yield rank-one matrices as their solutions; hence, the optimal beamforming schemes can be obtained by principal eigenvector decomposition of the SDP solutions.

As wireless communication systems evolve, besides the SINR constraints, a variety of additional constraints need to be considered for beamforming designs in diverse scenarios \cite{Zhang10, Zhang11,He14,Hua13,Huang10,Huang10b,Li13,Ham06}. In spectrum-sharing cognitive radio \cite{Zhang10,Zhang11} or heterogeneous networks \cite{He14}, the interference generated towards the co-channel users of other coexisting systems should be restricted or nulled. For secrecy communications, it is necessary to reduce the SINRs accessed by the eavesdroppers below prescribed thresholds \cite{Li13}; and in the emerging simultaneous wireless information and power transfer (SWIPT) scenarios, the power directed to the energy harvesting (EH) terminals should meet the specified targets \cite{Hua13}. These additional constraints can be modeled as {\em joint shaping constraints} in beamforming problems. Furthermore, to render desired beampattern per user, the works in \cite{Ham06,Huang10,Huang10b} considered the optimal beamforming designs with {\em individual shaping constraints}. With addition of such joint and individual shaping constraints, the SDP relaxation technique can be still applied for beamforming problem\cite{Luo10}. However, the resultant SDP is not guaranteed to yield rank-one solutions; in this case, SDP relaxation only provides a (probably unachievable) lower bound on the transmission power for linear beamforming schemes.

To approach the SDP bound, orthogonal space-time block codes (OSTBCs) were recently proposed to combine with the transmit-beamforming, in order to extend the conventional ``rank-one'' beamforming to a generalized ``rank-$K$'' beamforming while maintaining a low-complexity symbol detection \cite{Wen12,Sch15,Wu13,Law13,Law15,Tal15}. Specifically, Alamouti code based beamforming schemes were developed to enhance the multicasting or relaying performance in \cite{Wen12,Wu13, Sch15}, where Alamouti codeword is employed at the base station (BS) to allow two beamformers dedicated for symbols of each user in a low-complexity space-time manner. Such ``rank-two'' beamforming schemes were further extended to ``rank-$K$'' beamforming with high-order complex-valued \cite{Law13}, or real-valued OSTBCs \cite{Law15}. A general-rank multicasting beamforming strategy combined with space-time trellis coding (STTC) was also proposed in \cite{Tal15}. The orthogonality property of the OSTBC facilitates a low-complexity transceiver structure of the resultant rank-$K$ beamforming scheme, and leads to a same SDP relaxation as with the conventional rank-one beamforming problem. By allowing multiple beamformers to be optimized per user, an optimal OSTBC based beamforming scheme can be obtained when the SDP admits a solution of rank not larger than the OSTBC order; substantial performance gain over the rank-one beamforming schemes could be then available. However, the order of OSTBC cannot be flexibly selected; and more importantly, full-rate high-order OSTBC does not exist \cite{Tar99}. Due to the transmission rate loss, the SDP bound becomes unachievable in general for OSTBC based beamforming schemes. %{\color{blue} It is worth noting that, using the real-valued OSTBC, \cite{Law15} proposed rank-$K$ beamforming designs with full rate guarantees when $K\in{2,4,8}$. When the SDP solution rank is not greater than eight (after rank reduction procedure), the optimality is guaranteed for the design in \cite{Law15}. Otherwise, it fails to achieve the SDP bound.}

In this paper, we propose a novel REdundant-signal Embedded Linear BeamForming (REEL-BF) scheme with the following components:
\begin{itemize}
\item A redundant-signal embedded transmission structure, where one beamformer is devoted to information symbol and additional $K-1$ ``shaping beamformers'' are used for randomly generated $K-1$ redundant symbols per user.
\item Beamforming design with orthogonality constraints, where all $K-1$ shaping beamformers per user are restricted to be in the null space of (i.e., be orthogonal to) the downlink channel from the BS to this user.
\end{itemize}
As with the OSTBC based schemes\cite{Wen12,Sch15,Wu13,Law13,Law15}, the proposed redundant-signal embedded transmission structure together with the orthogonality constraints on shaping beamformers can enable an enlarged (i.e., rank-$K$) beamforming design space. Compared to the latter, a key difference is that the space-time precoder is not required. As a result, the information transmission and detection are performed on a symbol-by-symbol basis, and full-rate transmission is always ensured for arbitrary $K$ value. Surprisingly, we show that the resultant orthogonality-constrained beamforming design is sufficient to deliver an optimal linear beamforming solution that minimizes the BS transmission power under arbitrary shaping constraints. Specifically, we prove that our problem can be relaxed to an SDP, which is an equivalent ``unitarily rotated'' version of the SDP relaxation for classic (rank-one) beamforming problem. Capitalizing on this equivalence, we establish that the proposed REEL-BF scheme is capable of achieving the SDP transmission power bound when a sufficiently large $K$ value is selected in our design (i.e., when a sufficiently large number of redundant signals are generated to enlarge the beamforming design space). In particular, for any given downlink beamforming problem with arbitrary quadratic shaping constraints, we show that the value of $K$ can be pre-determined in accordance with the number of shaping constraints and/or the number of BS antennas, to ensure the optimality of the proposed REEL-BF design. Based on this result, an efficient algorithm is then developed to obtain the optimal REEL-BF solution that is guaranteed to achieve the SDP bound.

It is worth noting that the proposed REEL-BF approach is actually in a similar spirit with the ``energy-signal embedded'' SWIPT beamforming \cite{Xu14} and ``artificial noise (AN) aided'' physical-layer secrecy transmission schemes \cite{Zhou10,Liao11,LiMa13}.\footnote{ We would like to thank an anonymous reviewer for pointing out this connection as well as an independently developed work \cite{Li16} which addresses the link between the ``energy-signal embedded'' SWIPT and AN-aided secure communication methods.} In fact, our approach can be seen as a generalization of those methods in some sense, as it provides a systematic design framework and a low-complexity algorithm to find the optimal linear beamforming design under an arbitrary number of shaping constraints in the broad contexts of cognitive radio, heterogeneous, physical-layer security, and EH networks.

The remainder of the paper is organized as follows. Section II provides the channel model, and a brief review of the existing beamforming approaches. Section III describes the principles and key components of the proposed REEL-BF design, and develops an important alternative problem formulation. Section IV derives the sufficient conditions that ensure the optimality of the REEL-BF design, and proposes an efficient algorithm to obtain the optimal REEL-BF solution. Numerical results are provided to demonstrate the merit of the proposed scheme in Section V, followed by the conclusions.

{\em Notations}: The operations $(\cdot)^*$, $|\cdot|$, $(\cdot)^T$, $(\cdot)^H$, and $\|\cdot\|$ denote the conjugate, the absolute value of a scalar, the transpose, the conjugate transpose, and the Euclidean norm of a vector, respectively; $\boldsymbol{I}_n$ is the $n\times n$ identity matrix; $\boldsymbol{0}_{m\times n}$ is the $m\times n$ zero matrix; $\mathbb{E}[\cdot]$, $\text{tr}(\cdot)$, and $\text{rank}(\cdot)$ denote the statistical expectation, the matrix trace, and the matrix rank, respectively; $[\boldsymbol{a}]_n$ and $[\boldsymbol{A}]_{m,n}$ denote the $n$-th entry for vector $\boldsymbol{a}$ and the entry with row $m$ and column $n$ for matrix $\boldsymbol{A}$, respectively; $\boldsymbol{A}\succeq \boldsymbol{0}$ means that matrix $\boldsymbol{A}$ is positive semidefinite; $\boldsymbol{A}\bullet\boldsymbol{B}=\text{tr}(\boldsymbol{A}\boldsymbol{B})$ is the inner product between matrices $\boldsymbol{A}$, $\boldsymbol{B}$; $\text{diag}(\boldsymbol{a})$ denotes a diagonal matrix with the main diagonal given by $\boldsymbol{a}$; ${\cal CN}(\mu,\sigma^2)$ stands for the complex Gaussian variable with mean $\mu$ and variance $\sigma^2$.
%Let $\mathbb{H}^n_+$ be $n\times n$ positive semidefinite Hermitian matrix space.
%%%%%%%%%%%%%%%%%%%%%%%%%%%%%%%%%%%%%%%%%%%%%%%%%%%%%%%%%%%%%%%%%%%%%%
%                                                                    %
%               Section II                                %
%                                                                    %
%%%%%%%%%%%%%%%%%%%%%%%%%%%%%%%%%%%%%%%%%%%%%%%%%%%%%%%%%%%%%%%%%%%%%%

\section{Preliminaries}

\subsection{Channel Model}

Consider a MU-MISO downlink system where an $N_t$-antenna BS transmits independent signals to $M$ single-antenna users over a common frequency band simultaneously. The channel of each user is assumed to be frequency flat and remains constant during one data frame. Let $\boldsymbol{h}^H_m$ be an $N_t$-dimensional row vector representing the channel from the BS to user $m$. The baseband model of the received narrowband signal of user $m$ at time slot $n$ is
\begin{align}\label{eq.model}
 y_m(n) = \sum_{j=1}^M\boldsymbol{h}_m^H\boldsymbol{x}_j(n)+v_m(n), ~~ m=1,...,M,
\end{align}
where $\boldsymbol{x}_m(n)\in\mathbb{C}^{N_t\times 1}$ is the transmit signal vector intended for user $m$, $v_m(n)\sim{\cal CN}(0,\sigma_m^2)$ is the additive white Gaussian noise (AWGN) in time slot $n$, $\forall n=1,...,N$, and $N$ is the length of data frame. The transmission covariance matrix for user $m$ is $\mathbb{E}\left[ \boldsymbol{x}_m(n) \boldsymbol{x}^H_m(n)\right]$, $\forall m$, and the average transmission power per time slot is given by
%\begin{equation}
$P_t=\sum_{m=1}^M \text{tr}\left(\mathbb{E}\left[ \boldsymbol{x}_m(n) \boldsymbol{x}^H_m(n)\right]\right)$.
%\end{equation}

\subsection{Rank-One Beamforming Design}
Using the conventional (i.e., rank-one) beamforming strategy, the transmit signal for each user can be modeled as\cite{Ben01}:
\begin{align}\label{eq.rank-1}
\boldsymbol{x}_m(n)=\boldsymbol{w}_m s_m(n),~~m=1,...,M,
\end{align}
where $\boldsymbol{w}_m\in\mathbb{C}^{N_t\times 1}$ and $s_m(n)\in \mathbb{C}$ are the beamforming vector and the information symbol with unit power (i.e., $\mathbb{E}\left[|s_m(n)|^2\right]=1$) for user $m$, respectively. Then the average transmission power per time slot is given by
\begin{equation}
P_t=\sum_{m=1}^M \text{tr}(\boldsymbol{w}_m\boldsymbol{w}^H_m).
\end{equation}
Correspondingly, the receive signal model (\ref{eq.model}) becomes
\begin{align}\label{eq.rank-one}
 y_m(n) = \sum_{j=1}^M\boldsymbol{h}_m^H\boldsymbol{w}_j s_j(n)+v_m(n), ~~m=1,...,M.
\end{align}
Based on (\ref{eq.rank-one}), the SINR for user $m$ over one data frame can be calculated as
\begin{align}\label{eq.rank_one_sinr}
\text{SINR}_m=\frac{\boldsymbol{h}_m\boldsymbol{h}_m^H\bullet\boldsymbol{w}_m\boldsymbol{w}_m^H}{\sum_{j=1,j\neq m}^M\boldsymbol{h}_m\boldsymbol{h}_m^H\bullet\boldsymbol{w}_j\boldsymbol{w}_j^H+\sigma_m^2}, ~~\forall m.
\end{align}
Let $\gamma_m>0$ denote the minimum SINR requirement per user. The SINR constraint for user $m$ is then
\begin{align}\label{eq.sinr}
\text{SINR}_m\geq \gamma_m, ~~m=1,...,M,
\end{align}
which can be equivalently reformulated as
\begin{equation}\label{eq.sinr_recast}
\sum_{j=1}^M\boldsymbol{A}_{mj}\bullet \boldsymbol{w}_j\boldsymbol{w}_j^H \geq \sigma^2_m,~~ m=1,...,M,
\end{equation}
with, $\forall m,j$,\footnote{Note that we actually have $\boldsymbol{A}_{mj}=\boldsymbol{A}_{mk}$, $\forall j,k\neq m$; i.e., simplified notations can be used. Here we use $\boldsymbol{A}_{ij}$ for convenience and generality.}
\begin{equation}\label{eq.sinr_A}
\begin{cases}
\boldsymbol{A}_{mj} =\frac{1}{\gamma_m}\boldsymbol{h}_m\boldsymbol{h}_m^H,~~ j=m\\
\boldsymbol{A}_{mj} =-\boldsymbol{h}_m\boldsymbol{h}_m^H, ~\;~~j\neq m.
\end{cases}
\end{equation}

Besides the SINR constraints, it is desirable to incorporate additional shaping constraints for beamforming design in various MU-MISO scenarios. Essentially, there are two classes of shaping constraints, as detailed below.

\subsubsection{Joint Shaping Constraints}
In many applications, e.g., in the contexts of cognitive radio \cite{Zhang11}, heterogeneous \cite{He14}, physical-layer secrecy\cite{Li13}, and SWIPT networks \cite{Hua13}, it is important to control the amount of power generated along some particular directions. The general form of these joint shaping constraints can be formulated as \cite{Huang10}:
\begin{equation}\label{eq.soft}
 \sum_{j=1}^M \boldsymbol{A}_{ij} \bullet \boldsymbol{w}_j \boldsymbol{w}^H_j \unrhd_i \tau_i, ~~i=M+1,...,M+L,
\end{equation}
where Hermitian matrices $\boldsymbol{A}_{ij}$ and $\unrhd_i \in \{\leq,=,\geq\}$ are determined by specific applications, with corresponding thresholds $\tau_i$, $\forall i=M+1,...,M+L$, $j=1,...,M$. %Note that there is no mandatory positive semidefinite requirement for $\boldsymbol{A}_{ij}$, $\forall i,j$.

\subsubsection{Individual Shaping Constraints}
In general, we can consider the following $P$ groups of individual shaping constraints on the beamforming vectors (see\cite{Ham06,Huang10,Huang10b}):
\begin{equation} \label{eq.individual}
\begin{split}
\boldsymbol{C}_{pm}\bullet\boldsymbol{w}_m\boldsymbol{w}_m^H = 0, ~~\forall m\in {\cal E}_p\\
\boldsymbol{C}_{pm}\bullet\boldsymbol{w}_m\boldsymbol{w}_m^H \geq 0, ~~\forall m\in {\cal \bar{E}}_p,
\end{split}
\end{equation}
where ${\cal E}_p$ is the subset of the index set $\{1,...,M\}$, ${\cal \bar{E}}_p$ is the complement of ${\cal E}_p$, and $\boldsymbol{C}_{pm}$ can be any Hermitian matrices for $p=1,...,P$, $\forall m$. By properly selecting $\boldsymbol{C}_{pm}$, (\ref{eq.individual}) can lead to desired beam-pattern for each use \cite{Ham06, Huang10,Huang10b,Huang14}. General compact forms for (\ref{eq.individual}) can be written as
\begin{align}{\label{eq.double-sided}}
\ell_{pm}\leq \boldsymbol{C}_{pm}\bullet \boldsymbol{w}_m\boldsymbol{w}^H_m \leq \mu_{pm},~~ \forall p,~m,
 \end{align}
 where $\ell_{pm}$ and $\mu_{pm}$ are prescribed parameters with $\ell_{pm}\leq 0 \leq \mu_{pm}$, $p=1,...,P$, $m=1,...,M$.

\subsection{Conventional Optimal Beamforming Problem}

Assume that the perfect channel state information (CSI) $\{\boldsymbol{h}_m\}$ is available at the BS. We consider a typical downlink beamforming design that minimizes the transmission power subject to the SINR and additional joint/individual shaping constraints. Mathematically, the optimal rank-one beamforming problem can be formulated as\cite{Ger10}:
\begin{subequations}\label{eq.qp0}
\begin{align}
&\min_{\{\boldsymbol{w}_m \}}\,\sum_{m=1}^M \text{tr}(\boldsymbol{w}_m\boldsymbol{w}^H_m)\\
&\text{s.t.} \,\sum_{j=1}^M \boldsymbol{A}_{ij} \bullet \boldsymbol{w}_j \boldsymbol{w}_j^H \unrhd_i \tau_i,~~i=1,...,M+L,\\
& ~~~\ell_{pm}\leq\boldsymbol{C}_{pm} \bullet \boldsymbol{w}_{m}\boldsymbol{w}^H_m \leq \mu_{pm}, ~~p=1,...,P, ~~\forall m,
\end{align}
\end{subequations}
which is non-convex in general \cite{Huang10}. Define $\boldsymbol{X}_m:=\boldsymbol{w}_m\boldsymbol{w}_m^H$, $\forall m$. It is clear that
\begin{equation}
    \boldsymbol{X}_m \succeq \boldsymbol{0}, ~~ \text{rank}(\boldsymbol{X}_m)=1,~~ m=1,...,M.
\end{equation}
Dropping the rank-one constraints, the well-known SDP relaxation for (\ref{eq.qp0}) is:
\begin{subequations}\label{eq.sdp0}
\begin{align}
&\min_{\{\boldsymbol{X}_m \}}\,\sum_{m=1}^M \text{tr}(\boldsymbol{X}_m)\\
&\text{s.t.} \,\sum_{j=1}^M \boldsymbol{A}_{ij}\bullet \boldsymbol{X}_j \unrhd_i \tau_i,~~i=1,...,M+L\\
& ~~~~\ell_{pm}\leq\boldsymbol{C}_{pm}\bullet\boldsymbol{X}_{m}\leq \mu_{pm}, ~~p=1,...,P, ~~\forall m\\
& ~~~~\boldsymbol{X}_m\succeq \boldsymbol{0}, ~~ m=1,...,M,
\end{align}
\end{subequations}
where the first $M$ linear inequalities in (\ref{eq.sdp0}b) represent the SINR constraints, i.e., for $i=1,...,M$, $\boldsymbol{A}_{ij}$ are defined as in (\ref{eq.sinr_A}), $\tau_i:=\sigma_i^2$, and all $\unrhd_i$ are $\geq$; while $\boldsymbol{A}_{ij}$, $\forall i=M+1,...,M+L$, and $\boldsymbol{C}_{pm}$, $\forall p, m$, are appropriately chosen Hermitian matrices (not necessarily positive semidefinite). Note that (\ref{eq.sdp0}) is an instance of SDP, which can be efficiently solved by interior-points methods \cite{boyd2004}.\footnote{Throughout the paper, we assume that (\ref{eq.sdp0}) and its Lagrangian dual problem are both solvable; i.e., they have non-empty feasible sets. As the SDP (\ref{eq.sdp0}) is a convex problem, its infeasibility can be also determined by the available solvers. In the case of infeasibility, an admission control can be invoked to drop some users to render the problem solvable.}

The SDP (\ref{eq.sdp0}) can yield an optimal beamforming solution for (\ref{eq.qp0}) when it admits an optimal solution $\{\boldsymbol{X}^{\star}_m\}$, with $\text{rank}(\boldsymbol{X}^{\star}_m)=1$, $\forall m$. It has been proven that this optimality holds for certain cases, for example, when there are SINR constraints only (i.e., $L=P=0$) \cite{Ben01,Bjo14,Ger10}, or when the number of shaping constraints satisfies: $L=0$, $P\leq 2$ or $L\leq 2$, $P=0$~\cite{Huang10}, or when $P=0$, $\boldsymbol{A}_{ij}\succeq \boldsymbol{0}$ (they can be of any rank), and all $\unrhd_i$ are $\leq$, $\forall i=M+1,...,M+L$, $j=1,...,M$~\cite{Zhang10}. In general, however, there are no guarantees that (\ref{eq.sdp0}) always admits a rank-one solution. Hence, the solution of (\ref{eq.sdp0}) only provides a lower bound for the transmission power in (\ref{eq.qp0}). In the case of $\text{rank}(\boldsymbol{X}^{\star}_m)>1$ for some $m$, a Gaussian randomization procedure can be employed to obtain an approximate (suboptimal) solution \cite{Luo10,Law15}.

\subsection{OSTBC based Beamforming Design}

For convenience, we call the transmission power bound promised by the solution of (\ref{eq.sdp0}) the SDP bound hereinafter. To approach such an SDP bound, an intuitive idea is to allow multiple beamformers assigned for information symbols of each user. To this end, a few recent works \cite{Wen12,Sch15,Wu13,Law13,Law15} proposed to combine OSTBC and downlink beamforming design at the BS. For the purpose of our discussion, let us start with a revisit of the rank-two beamforming scheme enabled by Alamouti code in \cite{Wen12,Sch15,Wu13,Law13}.

Let the information symbol stream $s_m(n)$ per user $m$ be grouped into blocks of two symbols; i.e., $\boldsymbol{s}_m(q)=[s_m(2q-1),\,s_m(2q)]^T$, $q=1,...,N/2$. For the ease of presentation, we set $q=1$ and then drop the block index of $\boldsymbol{s}_m$ without loss of generality; i.e., we focus on $\boldsymbol{s}_m=[s_m(1),\,s_m(2)]^T$. Combining the beamforming scheme with Alamouti code at the BS, the transmit space-time block is
\begin{align}
\left[\boldsymbol{x}_1,\,\boldsymbol{x}_2 \right]=\sum_{m=1}^M\left[\boldsymbol{w}_{m,1}, \,\boldsymbol{w}_{m,2}\right]{\cal C}(\boldsymbol{s}_m),
\end{align}
where ${\cal C}(\cdot)$ is the Alamouti-coding matrix:
\begin{equation}
{\cal C}(\boldsymbol{s}_m):=
\begin{bmatrix}
s_m(1) & -s_m^*(2)\\
s_m(2) & s_m^*(1)\\
\end{bmatrix},
\end{equation}
and $\boldsymbol{w}_{m,1}$ and $\boldsymbol{w}_{m,2}$ are two beamformers assigned for user $m$. With $\boldsymbol{W}_m:=[\boldsymbol{w}_{m,1},\,\boldsymbol{w}_{m,2}]$, $\forall m$, the transmission covariance matrix for user $m$ can be calculated as $\boldsymbol{W}_m\boldsymbol{W}^H_m$, $\forall m$, and the average transmission power per time slot is
\begin{equation}
P_t=\sum_{m=1}^M \sum_{k=1}^2 \|\boldsymbol{w}_{m,k}\|^2=\sum_{m=1}^M \text{tr}(\boldsymbol{W}_m\boldsymbol{W}^H_m).
\end{equation}

Using the orthogonality property of ${\cal C}(\cdot)$, the equivalent single-input single-output (SISO) model is obtained\cite{Tar99}:
\begin{equation*}\label{eq.Ala}
\begin{split}
\begin{bmatrix}
\tilde{s}_{m}(1)\\\tilde{s}_{m}(2)
\end{bmatrix} =
\sum_{j=1}^M
\sqrt{\boldsymbol{h}_m\boldsymbol{h}_m^H\bullet\boldsymbol{W}_j\boldsymbol{W}_j^H }
\begin{bmatrix}
s_{j}(1)\\s_{j}(2)
\end{bmatrix}
+&\begin{bmatrix}
\tilde{v}_m(1)\\\tilde{v}_m(2)
\end{bmatrix},
\end{split}
\end{equation*}
where $\tilde{v}_m(1)$, $\tilde{v}_m(2)\sim {\cal CN}\left(0, \sigma_m^2\right)$. Each symbol can be then independently detected, and the SINR for retrieving $s_m(1)$ or $s_m(2)$ is characterized by
\begin{equation}\label{eq.OSTBC_sinr}
\begin{split}
\text{SINR}_m
&=\frac{\boldsymbol{h}_m\boldsymbol{h}_m^H\bullet\boldsymbol{W}_m\boldsymbol{W}_m^H}{\sum_{j\neq m} \boldsymbol{h}_m\boldsymbol{h}_m^H\bullet\boldsymbol{W}_j\boldsymbol{W}_j^H+\sigma_m^2},~~\forall m.
\end{split}
\end{equation}
Note that (\ref{eq.OSTBC_sinr}) becomes identical to (\ref{eq.rank_one_sinr}) when $\boldsymbol{W}_m$ is substituted with $\boldsymbol{w}_m$ for $m=1,...,M$. In addition, the joint and individual shaping constraints become identical to (\ref{eq.soft}) and (\ref{eq.double-sided}) with $\boldsymbol{w}_m$ replaced by $\boldsymbol{W}_{m}$, $\forall m$.

Let $\boldsymbol{X}_m:=\boldsymbol{W}_m\boldsymbol{W}_m^H$, $m=1,...,M$. It follows that
\begin{equation}
\boldsymbol{X}_m\succeq \boldsymbol{0},~~ \text{rank}(\boldsymbol{X}_m)\leq 2,~~ \forall m.
\end{equation}
Removing the rank-two constraints, one can obtain the {\em same} formulation as (\ref{eq.sdp0}) for the Alamouti-code assisted beamforming design. Clearly, with the help of the Alamouti code, the proposed beamforming design admits a larger feasible region than the rank-one beamforming design. For the optimal solution $\{\boldsymbol{X}^{\star}_m\}$ of (\ref{eq.sdp0}), as long as $\text{rank}(\boldsymbol{X}^{\star}_m)\leq 2$, $\forall m$, the optimal beamformers can be retrieved by proper matrix decompositions. Note that this is achieved without loss of transmission rate (also known as bandwidth efficiency), and without significantly increased complexity in transceiver implementations. The existence of rank-two optimal solutions for (\ref{eq.sdp0}) only holds under conditions, e.g., when the number of shaping constraints satisfies: $L+PM<8$ \cite{Huang10}. When (\ref{eq.sdp0}) admits a solution with $\text{rank}(\boldsymbol{X}^{\star}_m) > 2$ for a certain $m$, the Alamouti-code based scheme fails to achieve the SDP bound, and a Gaussian randomization procedure is called to compute an approximate solution based on $\{\boldsymbol{X}^{\star}_m\}$ \cite{Wu13}.

The Alamouti-code based scheme can be generalized for the high-order OSTBC. In particular, using the real-valued OSTBC, \cite{Law15} proposed rank-$K$ beamforming designs with full-rate guarantees where $K \in \{2,4,8\}$. However, it is well known that full-rate high-order (i.e., $K> 2$ for complex-valued, and $K>8$ for real-valued) OSTBC does not always exist. Hence, when (\ref{eq.sdp0}) requires high-rank solutions, the SDP bound cannot be achieved by either beamforming design with full-rate low-order OSTBC, or that with the high-order OSTBC since the SINR targets need to be adjusted to compensate the rate loss due to the OSTBC in use. This is also demonstrated by numerical results in Sec.~V.

%Recall that the OSTBC was originally proposed to achieve the transmit-diversity without CSI at the transmitter; yet, the CSI is actually available for beamforming design here. This motivates our optimal space-time beamforming design to achieve the performance of the relaxed (\ref{eq.sdp0}) in the sequel.

%%%%%%%%%%%%%%%%%%%%%%%%%%%%%%%%%%%%%%%%%%%%%%%%%%%%
%                                                                                  %
%               Section: Channel-aware Beamforming Design       %
%                                                                                  %
%%%%%%%%%%%%%%%%%%%%%%%%%%%%%%%%%%%%%%%%%%%%%%%%%%%%%

\section{Redundant-Signal Embedded Linear BeamForming Design}

The OSTBC based schemes introduce the idea of developing rank-$K$ beamforming in a space-time manner. While the orthogonality of OSTBC facilitates the low-complexity transceiver structure, this requirement also leads to the loss of transmission rate such that the SDP bound cannot be achieved in general. In this section, we propose a novel REEL-BF scheme where a ``redundant-signal embedded'' transmission strategy is employed to create an enlarged (i.e., general rank-$K$) beamforming design space at the BS. 

\subsection{Redundant-Signal Embedded Transmission Strategy}

To implement the proposed redundant-signal embedded transmission strategy, in addition to the information-bearing signal $s_m$, we generate $K-1$ independent and identically distributed (i.i.d.) random signals $z_{m,k}$, $k=2,...,K$, per user $m$. Without loss of generality, we assume that $s_m$ and $z_{m,k}$ are unit power; i.e., $\mathbb{E}[|s_m|^2]=1$, $\forall m$, and $\mathbb{E}[|z_{m,k}|^2]=1$, $k=2,...,K$, $\forall m$.  Let $\boldsymbol{w}_{m,1}\in\mathbb{C}^{N_t\times 1}$ denote the beamformer devoted to the information signal $s_m$, and $\boldsymbol{w}_{m,k}\in\mathbb{C}^{N_t\times 1}$ the beamformers used for ``redundant'' signals $z_{m,k}$, $k=2,...,K$. The transmitted signal for user $m$ is then constructed as
\begin{equation}\label{eq.prop_model0}
\boldsymbol{x}_m = \boldsymbol{w}_{m,1}s_m +  \sum_{k=2}^K \boldsymbol{w}_{m,k}z_{m,k}.
\end{equation}
Here, the redundant signals $z_{m,k}$ are in fact generated to produce the desirable radiation beam-pattern for e.g. energy harvesting \cite{Xu14} or general co-channel interference control purposes \cite{Law15}, such that the additional shaping constraints can be met in the intended beamforming design. For this reason, we call $\boldsymbol{w}_{m,k}$, $k=2,...,K$, the {\em shaping} beamformers. Note that these beamformers are not used to shape the main information-bearing beamformer, but to ``shape'' the overall radiation beam-pattern at the BS in accordance with the additional joint/individual shaping constraints.

We would like to emphasize the following fact:
\begin{fact}
The value of $K$ can be flexibly chosen as an arbitrary positive integer in the proposed REEL-BF design.
\end{fact}
The freedom in selection of the value $K$ will play an important role in ensuring the optimality of our design, as will be shown in the sequel.

Let $\boldsymbol{W}_m:=[\boldsymbol{w}_{m,1},...,\boldsymbol{w}_{m,K}]$, $\forall m$. The baseband transmit signal at the BS is given by
\begin{equation}\label{eq.prop_model}
\boldsymbol{x} = \sum_{m=1}^M \boldsymbol{x}_m = \sum_{m=1}^M \boldsymbol{W}_{m}[s_m,z_{m,2},...,z_{m,K}]^T.
\end{equation}
Based on (\ref{eq.prop_model}), the average transmission power at the BS is
\begin{equation}\label{eq.TxPower}
P_t=\sum_{m=1}^M \sum_{k=1}^K \|\boldsymbol{w}_{m,k}\|^2= \sum_{m=1}^M \text{tr}( \boldsymbol{W}_{m}\boldsymbol{W}_{m}^H ).
\end{equation}

Given the downlink channel vector $\boldsymbol{h}^H_m$, the received signal at user $m$ is
\begin{equation}\label{eq.model1}
\begin{split}
 y_m  = & ~\boldsymbol{h}_m^H \boldsymbol{w}_{m,1}s_m + \sum_{k=2}^K \boldsymbol{h}_m^H\boldsymbol{w}_{m,k} z_{m,k} \\
 &+\sum_{j=1,j\neq m}^M \boldsymbol{h}_m^H\left(\boldsymbol{w}_{j,1}s_{j}+\sum_{k=2}^K \boldsymbol{w}_{j,k}z_{j,k}\right) + v_m,
 \end{split}
\end{equation}
where ${v}_m\sim\mathcal{CN}(0,\sigma^2_m)$ denotes the AWGN. The SINR for user $m$ is then
\begin{equation}\label{eq.sinr_reel}
    \text{SINR}_{m} =\frac{|\boldsymbol{h}^H_m\boldsymbol{w}_{m,1}|^2}{\sum_{k=2}^K |\boldsymbol{h}^H_m\boldsymbol{w}_{m,k}|^2+\sum_{j\neq m} \sum_{k=1}^K |\boldsymbol{h}^H_m\boldsymbol{w}_{j,k}|^2+\sigma_m^2}.
\end{equation}

\subsection{Beamforming Design with Orthogonality Constraints}

At the first sight, the proposed redundant signal embedded strategy seems ill-conceived, as the randomly generated redundant signals are not useful for information transmission and can even produce the interference to the same users' information symbols. To eliminate the latter interference, we impose the following orthogonality constraints on the shaping beamformers $\{\boldsymbol{w}_{m,2},\ldots, \boldsymbol{w}_{m,K}\}$:
\begin{equation}\label{eq.ortho}
\boldsymbol{h}^H_m\boldsymbol{w}_{m,k}= 0,~~ k=2,...,K,
\end{equation}
for $m = 1, \ldots, M$. Namely, we allow only the information beamformer, $\boldsymbol{w}_{m,1}$, to be optimized in whole space while all the $K-1$ shaping beamformers, i.e., $\boldsymbol{w}_{m,2},...,\boldsymbol{w}_{m,K}$, are required to stay in the null space of $\boldsymbol{h}_m$ per user $m$. Somewhat surprisingly, it will be shown that such an orthogonality-constrained design is sufficient to deliver an optimal linear beamforming scheme that achieves the SDP bound.

With the constraints (\ref{eq.ortho}), the SINR in (\ref{eq.sinr_reel}) simplifies to
\begin{equation}\label{eq.lem_SINR}
\begin{split}
\text{SINR}_{m}
&=\frac{|\boldsymbol{h}^H_m\boldsymbol{w}_{m,1}|^2}{\sum_{j\neq m} \sum_{k=1}^K |\boldsymbol{h}^H_m\boldsymbol{w}_{j,k}|^2+\sigma_m^2}\\
&=\frac{\boldsymbol{h}_m\boldsymbol{h}^H_m\bullet \boldsymbol{W}_m\boldsymbol{W}_m^H}{\sum_{j\neq m} \boldsymbol{h}_m\boldsymbol{h}^H_m\bullet \boldsymbol{W}_{j}\boldsymbol{W}_j^H+\sigma_m^2},~~\forall m,\\
\end{split}
\end{equation}
where the second equality holds since $\boldsymbol{h}_m\boldsymbol{h}_m^H\bullet\boldsymbol{W}_j\boldsymbol{W}_j^H=\sum_{k=1}^K |\boldsymbol{h}^H_m\boldsymbol{w}_{j,k}|^2$ and $\boldsymbol{h}_m\boldsymbol{h}_m^H\bullet\boldsymbol{W}_m\boldsymbol{W}_m^H=|\boldsymbol{h}^H_m\boldsymbol{w}_{m,1}|^2$ due to $\boldsymbol{h}^H_m\boldsymbol{w}_{m,k}= 0$, $\forall k=2,...,K$. Clearly, the SINR expression reduces to the same form with (\ref{eq.OSTBC_sinr}).

The orthogonality constraints (\ref{eq.ortho}) can be rewritten in a compact matrix form:
\begin{equation}\label{eq.ortho1}
    \boldsymbol{D}\bullet \boldsymbol{W}_m^H \boldsymbol{h}_m \boldsymbol{h}_m^H \boldsymbol{W}_m=0, ~~ \forall m,
\end{equation}
where $\boldsymbol{D}:= \text{diag}(0,1,\ldots, 1)$ is a $K\times K$ diagonal matrix. Under the orthogonality constraints, the SINR constraints as well as other joint and individual shaping constraints become identical to (\ref{eq.sinr_recast}), (\ref{eq.soft}) and (\ref{eq.double-sided}) with $\boldsymbol{w}_m$ replaced by $\boldsymbol{W}_{m}$, $\forall m$. As a result, the intended power minimization problem can be formulated as:
\begin{subequations}\label{eq.sdp1}
\begin{align}
& \min_{\{\boldsymbol{W}_m\}}\,\sum_{m=1}^M \text{tr}(\boldsymbol{W}_m\boldsymbol{W}_m^H)\\
&\text{s.t.} \,\sum_{j=1}^M \boldsymbol{A}_{ij}\bullet \boldsymbol{W}_j\boldsymbol{W}_j^H \unrhd_i \tau_i,~~i=1,...,M+L\\
& ~~~\ell_{pm}\leq\boldsymbol{C}_{pm}\bullet\boldsymbol{W}_m\boldsymbol{W}_m^H\leq \mu_{pm}, ~p=1,...,P, ~\forall m \\
& ~~~\boldsymbol{W}_m\boldsymbol{W}_m^H\succeq \boldsymbol{0},~ \text{rank}(\boldsymbol{W}_m\boldsymbol{W}_m^H)\leq K,~~ \forall m \label{eq.rank-K}\\
& ~~~ \boldsymbol{D}\bullet \boldsymbol{W}_m^H \boldsymbol{h}_m \boldsymbol{h}_m^H \boldsymbol{W}_m=0, ~~ \forall m. \label{eq.ortho2}
\end{align}
\end{subequations}
Note that the constraints in (\ref{eq.rank-K}) are in fact redundant; we include them for an easy comparison with (\ref{eq.sdp0}).

\begin{remark}
It is clearly shown by (\ref{eq.sdp1}) that the proposed redundant-signal embedded structure together with the orthogonality constraints (\ref{eq.ortho}) enable a rank-$K$ beamforming design at the BS, as with the OSTBC-based approaches in \cite{Wen12,Sch15,Wu13,Law13,Law15}. Yet, compared to the latter, the proposed scheme has two significant differences: i) the information transmission and detection are performed on a symbol-by-symbol (instead of block-by-block) basis, hence, no (block) encoding and decoding delays are incurred; and ii) the full-rate transmission is always ensured, regardless of the choice of $K$ value. While the first feature is definitely valuable for practical (e.g., real-time) applications, the second one will be the key to overcome the limitation of the OSTBC-based approaches such that the SDP bound can be always achieved.
\end{remark}

%Note that the shaping beamformers cannot contribute to the signal of interest, even when each $z_{m,k}$ bears the information for user $m$. However, with careful design along with $\boldsymbol{w}_{m,1}$, these shaping beamformers are essentially useful for producing desirable radiation pattern at the BS to meet the general constraints under consideration. In fact, as will be demonstrated in the Sec. IV and corroborated by numerical results in Sec.~V, these orthogonal constraints incur no performance degradation. By performing a judicious basis rotation, one can obtain an equivalent SDP relaxation with that of classical beamforming design.

\subsection{Alternative Formulation}

Due to the presence of the orthogonality constraints (\ref{eq.ortho2}), the problem cannot be reduced to the SDP in (\ref{eq.sdp0}) by simply removing the rank constraints $\text{rank}(\boldsymbol{W}_m\boldsymbol{W}_m^H)\leq K$, $\forall m$. We next show how to deal with such orthogonality constraints in a simple ``unitary rotation'' manner, and perform a judicious change of optimization variables to obtain an alternative formulation for (\ref{eq.sdp1}), which will turn out to play an important role in computing the optimal REEL-BF solution.

To this end, we first define the normalized downlink channel for user $m$ by
\begin{equation}
 \boldsymbol{\bar{h}}_m:=\boldsymbol{h}_m/\|\boldsymbol{h}_m\|, ~~m=1,...,M.
\end{equation}
Let the columns of $\boldsymbol{F}_m\in\mathbb{C}^{N_t\times (N_t-1)}$ constitute an {\em orthonormal basis} for the null space of $\boldsymbol{h}_m\boldsymbol{h}_m^H$, denoted by
\begin{equation}
\boldsymbol{F}_m=\text{null}(\boldsymbol{\bar{h}}_m\boldsymbol{\bar{h}}_m^H), ~~m=1,...,M.
\end{equation}
It is ready to obtain $\boldsymbol{F}_m$ by the eigenvalue decomposition of $\boldsymbol{\bar h}_m\boldsymbol{\bar h}^H_m$. Specifically, we have
\begin{equation}\label{eq.U}
 \boldsymbol{\bar h}_m\boldsymbol{\bar h}_m^H = \boldsymbol{U}_m \boldsymbol{\Lambda} \boldsymbol{U}_m^H
= \left[ \boldsymbol{\bar{h}}_{m} \,\, \boldsymbol{F}_m \right]
\boldsymbol{\Lambda}
\begin{bmatrix} \boldsymbol{\bar{h}}^H_{m} \\ \boldsymbol{F}^H_m\end{bmatrix},~~\forall m,
\end{equation}
where $\boldsymbol{\Lambda}:=\text{diag}(1,0,...,0)$ is an $N_t \times N_t$ diagonal matrix. Note that $\boldsymbol{U}_m=[\boldsymbol{\bar{h}}_{m} \, \,\boldsymbol{F}_m]$, $\forall m$, are $N_t \times N_t$ unitary matrices. The columns of $\boldsymbol{U}_m$ can serve as an orthonormal basis for a ``rotated'' full space of $\mathbb{C}^{N_t\times 1}$.

Given $\boldsymbol{\bar h}_m$ and $\boldsymbol{F}_m$, $\forall m$, the orthogonality constraints in (\ref{eq.ortho}) actually imply that the beamforming vectors under consideration are
\begin{equation}\label{eq.beam_str1}
\left\{ \begin{split}
\boldsymbol{w}_{m,1}&=&&\alpha_m\boldsymbol{\bar h}_m+\boldsymbol{F}_m\boldsymbol{\beta}_m\\
[\boldsymbol{w}_{m,2},...,\boldsymbol{w}_{m,K}] &=&& \boldsymbol{F}_m \boldsymbol{\Omega}_m, %[\boldsymbol{\varpi}_{m,1},...,\boldsymbol{\varpi}_{m,K-1}],\\
\end{split}\right.
\end{equation}
where $\alpha_m\in \mathbb{R}$, $\boldsymbol{\beta}_m\in\mathbb{C}^{(N_t-1)\times 1}$, and $\boldsymbol{\Omega}_m\in\mathbb{C}^{(N_t-1)\times (K-1)}$, $\forall m$.\footnote{In general, $\alpha_m$ should be complex-valued, i.e., $\alpha_m \in \mathbb{C}$. However, since an arbitrary phase rotation for $\alpha_m$ would not affect both the transmission power and the quadratic constraints of interest, we can simply assume $\alpha_m\in\mathbb{R}$ without loss of optimality.} Clearly, the design of $\{\boldsymbol{w}_{m,k}\}$ is equivalent to determine the ``coordinates'' $\{\alpha_m, \boldsymbol{\beta}_m, \boldsymbol{\Omega}_m\}$ in the rotated space defined by $\boldsymbol{U}_m$, $\forall m$. These coordinates would be the variables to be optimized in our alternative problem formulation.

Based on (\ref{eq.U}) and (\ref{eq.beam_str1}), it readily follows that
\begin{align}\label{eq.beam_str2}
 \boldsymbol{W}_m=
\boldsymbol{U}_{m}
  \begin{bmatrix}
  \alpha_m& \boldsymbol{0}_{1\times (K-1)}\\
  \boldsymbol{\beta}_m & \boldsymbol{\Omega}_m
  \end{bmatrix}, ~~\forall m,
\end{align}
where the orthogonality constraints (\ref{eq.ortho}) are implicitly absorbed in this special structure. The transmission covariance matrix for user $m$ is then
\begin{equation}\label{eq.rotate_covariance}
\boldsymbol{X}_m= \boldsymbol{W}_m\boldsymbol{W}_m^H=\boldsymbol{U}_m\boldsymbol{\hat{X}}_m\boldsymbol{U}_m^H, ~~m=1,...,M,
\end{equation}
where
\begin{equation}\label{eq.covariance}
    \boldsymbol{\hat{X}}_m:=  \begin{bmatrix}
  \alpha_m^2& \alpha_m\boldsymbol{\beta}^H_m\\
  \alpha_m\boldsymbol{\beta}_m & \boldsymbol{\beta}_{m}\boldsymbol{\beta}^H_m+\boldsymbol{\Omega}_m\boldsymbol{\Omega}^H_m
  \end{bmatrix},~~\forall m.
\end{equation}
Clearly, $\boldsymbol{\hat X}_m \succeq \boldsymbol{0}$, $\forall m$. According to (\ref{eq.rotate_covariance}), the orthogonality constraints (\ref{eq.ortho}) simply require that covariance matrix $\boldsymbol{X}_m$ be structured to be unitarily similar to $\boldsymbol{\hat X}_m$ in (\ref{eq.covariance}). Interestingly, as will be demonstrated in Sec.~IV, such a structure incurs no performance loss due to two facts: i) unitary rotation is invertible, and ii) matrix $\boldsymbol{\hat X}_m$ can be constructed from any positive semidefinite matrix.

Based on (\ref{eq.rotate_covariance}), the average transmission power in (\ref{eq.TxPower}) is rewritten as
\begin{align}
 P_t &=\sum_{m=1}^M \text{tr}\left(\boldsymbol{U}_m\boldsymbol{\hat{X}}_m\boldsymbol{U}_m^H\right),
\end{align}
and the SINR constraints in (\ref{eq.lem_SINR}) can be reformulated as:
\begin{equation}\label{eq.sinr_pro}
\sum_{j=1}^M\boldsymbol{A}_{mj}\bullet \boldsymbol{U}_j\boldsymbol{\hat X}_j
   \boldsymbol{U}_j^H \geq \sigma^2_m,~~ m=1,...,M.
\end{equation}
Similarly, the context-specific joint/individual shaping constraints are in the same forms as those in (\ref{eq.soft}) or (\ref{eq.double-sided}) with $\boldsymbol{w}_m\boldsymbol{w}_m^H$ replaced by $\boldsymbol{U}_m\boldsymbol{\hat{X}}_m \boldsymbol{U}_m^H$, $\forall m$.

Hence, the proposed REEL-BF design problem (\ref{eq.sdp1}) can be alternatively formulated as:
\begin{subequations}\label{eq.prob_qp}
 \begin{align}
&\min_{\{\alpha_m, \boldsymbol{\beta}_{m},\boldsymbol{\Omega}_{m}\}} ~ \sum_{m=1}^M \text{tr}\left( \boldsymbol{U}_m\boldsymbol{\hat{X}}_m\boldsymbol{U}_m^H\right)\\
& \text{s.t.} \, \sum_{j=1}^M \boldsymbol{A}_{ij}\bullet
  \boldsymbol{U}_j\boldsymbol{\hat{X}}_j\boldsymbol{U}^H_j \unrhd_i \tau_i, ~~ i=1,....,M+L\\
&\ell_{pm}\leq\boldsymbol{C}_{pm} \bullet
  \boldsymbol{U}_m \boldsymbol{\hat{X}}_m\boldsymbol{U}^H_m \leq \mu_{pm},~p=1,...,P, ~\forall m\\
 & \boldsymbol{\hat{X}}_m:=  \begin{bmatrix}
  \alpha_m^2& \alpha_m\boldsymbol{\beta}^H_m\\
  \alpha_m\boldsymbol{\beta}_m & \boldsymbol{\beta}_{m}\boldsymbol{\beta}^H_m+\boldsymbol{\Omega}_m\boldsymbol{\Omega}^H_m
  \end{bmatrix} \succeq \boldsymbol{0},~ ~ \forall m. \label{eq.hat_X}
\end{align}
\end{subequations}
Note that the orthogonality constraints (\ref{eq.ortho2}) are guaranteed by both (\ref{eq.beam_str1}) and the definition of $\boldsymbol{U}_m$ in (\ref{eq.U}); hence, they can be omitted here. The problem (\ref{eq.prob_qp}) is non-convex due to the quadratic forms for entries in $\boldsymbol{\hat{X}}_m$, $\forall m$, and the coupling of the optimization variables. However, it can be readily relaxed to a convex SDP, as will be described next.

%In the next section, we will show that, with proper selection of $K$, (\ref{eq.prob_qp}) can achieve the performance of (\ref{eq.sdp0}), and the optimal solution to (\ref{eq.prob_qp}) can be efficiently obtained by solving (\ref{eq.sdp0}). .

\section{SDP based Approach to Optimal REEL-BF Solution}

In this section, we develop an efficient SDP based approach to solving the REEL-BF problem (\ref{eq.prob_qp}) (or equivalently, (\ref{eq.sdp1})), and show that our REEL-BF design can always achieve the SDP bound.

\subsection{SDP Relaxation of (\ref{eq.prob_qp}) and Its Equivalence to (\ref{eq.sdp0})}

Define new variables $\eta_m:=\alpha_m^2$, $\boldsymbol{\xi}_{m}:={\alpha}_m\boldsymbol{\beta}_{m}$, and $\boldsymbol{\Gamma}_{m}:=\boldsymbol{\beta}_{m}\boldsymbol{\beta}_{m}^H+\boldsymbol{\Omega}_m\boldsymbol{\Omega}^H_m$, $\forall m$. With such definitions, we readily have
\begin{equation*}
    \eta_m \boldsymbol{\Omega}_m\boldsymbol{\Omega}^H_m = \eta_m \boldsymbol{\Gamma}_m-\boldsymbol{\xi}_m\boldsymbol{\xi}_m^H,~~\forall m.
\end{equation*}
It follows that $\boldsymbol{\hat X}_m\succeq \boldsymbol{0}$, $\forall m$, are equivalent to
\begin{equation}\label{eq.pos_semi}
    \eta_m\geq 0, ~~\boldsymbol{\Gamma}_m\succeq \boldsymbol{0},~~ \eta_m \boldsymbol{\Gamma}_m-\boldsymbol{\xi}_m\boldsymbol{\xi}_m^H \succeq \boldsymbol{0}, ~~\forall m,
\end{equation}
and
\begin{equation}\label{eq.prob_rank}
\text{rank}(\eta_m \boldsymbol{\Gamma}_m-\boldsymbol{\xi}_m\boldsymbol{\xi}_m^H)\leq  K-1, ~~\forall m,
\end{equation}
where the rank constraints are simply due to $\boldsymbol{\Omega}_m\in\mathbb{C}^{(N_t-1)\times (K-1)}$, $\forall m$.

Based on the Schur complement\cite[A.5.5]{boyd2004}, the constraints in (\ref{eq.pos_semi}) are equivalent to\footnote{Note that this equivalence actually requires $\eta_m>0$, $\forall m$, which is always guaranteed for any given positive SINR targets, as will be shown in Lemma~2.}
\begin{equation} \label{eq.bar_X}
    \boldsymbol{\bar{X}}_m:=  \begin{bmatrix}
  \eta_m& \boldsymbol{\xi}^H_m\\
  \boldsymbol{\xi}_m & \boldsymbol{\Gamma}_m
  \end{bmatrix} \succeq \boldsymbol{0}, ~~\forall m.
\end{equation}

Removing the rank constraints in (\ref{eq.prob_rank}) and substituting $\{\boldsymbol{\hat X}_m\}$ with $\{\boldsymbol{\bar X}_m\}$ in (\ref{eq.prob_qp}), the SDP relaxation for (\ref{eq.prob_qp}) can be thus obtained as:
\begin{subequations} \label{eq.prob_sdp}
 \begin{align}
&\min_{\{ \boldsymbol{\bar X}_{m}\}} ~ \sum_{m=1}^M
\text{tr}(\boldsymbol{U}_m\boldsymbol{\bar{X}}_m\boldsymbol{U}_m^H)\\
& \text{s.t.} ~\sum_{j=1}^M \boldsymbol{A}_{ij}\bullet \boldsymbol{U}_j\boldsymbol{\bar X}_j\boldsymbol{U}^H_j
   \unrhd_i \tau_i, ~~ \forall i \\
& ~~~~~\ell_{pm}\leq \boldsymbol{C}_{pm}\bullet
  \boldsymbol{U}_m\boldsymbol{\bar X}_m\boldsymbol{U}^H_m
   \leq \mu_{pm},~~\forall p,~ m\\
  &~~~~~ \boldsymbol{\bar X}_m:=\begin{bmatrix}
  \eta_m& \boldsymbol{\xi}^H_m\\
  \boldsymbol{\xi}_m & \boldsymbol{\Gamma}_m
  \end{bmatrix}\succeq \boldsymbol{0},~~\forall m.
  \end{align}
  \end{subequations}
Note that if we simply ignore the orthogonality constraints (\ref{eq.ortho1}), remove the rank constraints, and set $\boldsymbol{X}_m =\boldsymbol{W}_m\boldsymbol{W}_m^H$, $\forall m$, then the problem (\ref{eq.sdp1}) can be directly relaxed to (\ref{eq.sdp0}). To facilitate our discussion, let us call (\ref{eq.prob_sdp}) the {\em rotated} SDP relaxation, and (\ref{eq.sdp0}) the {\em original} SDP relaxation. It appears that the rotated SDP relaxation (\ref{eq.prob_sdp}) could be tighter than the original SDP relaxation (\ref{eq.sdp0}) since it takes into account the orthogonality constraints. Interestingly, we show the equivalence between these two SDPs in the following lemma.

\begin{lemma}
The rotated SDP relaxation (\ref{eq.prob_sdp}) is equivalent to the original SDP relaxation (\ref{eq.sdp0}).
\end{lemma}

\begin{IEEEproof}
Given any feasible solution $\{\boldsymbol{\bar X}_m\}$ for (\ref{eq.prob_sdp}), it is clear that the positive semidefinite matrix set $\{\boldsymbol{X}_m\}$ with $\boldsymbol{X}_m=\boldsymbol{U}_m\boldsymbol{\bar X}_m\boldsymbol{U}^H_m$, $\forall m$, is also feasible for (\ref{eq.sdp0}), and achieves the same total transmission power value.

Similarly, given any feasible positive semidefinite matrix set $\{\boldsymbol{X}_m\}$ for (\ref{eq.sdp0}), we can construct the positive semidefinite matrices $\boldsymbol{\bar X}_m=\boldsymbol{U}_m^H\boldsymbol{X}_m\boldsymbol{U}_m$, $\forall m$, which are feasible for (\ref{eq.prob_sdp}). The equivalence readily follows.
\end{IEEEproof}

\begin{remark}
Lemma 1 provides us an interesting insight. Although the orthogonality constraints (\ref{eq.ortho1}) (or equivalently, (\ref{eq.ortho})) are accommodated by the proposed REEL-BF scheme, the feasible set for the rotated SDP relaxation (\ref{eq.prob_sdp}) is isomorphic (with respect to unitary rotation) to that for the original SDP relaxation (\ref{eq.sdp0}). Namely, the additional orthogonality constraints (\ref{eq.ortho1}) only result in a unitary rotation from any feasible transmission covariance matrix $\boldsymbol{X}_m$ for (\ref{eq.sdp0}) to a feasible $\boldsymbol{\bar X}_m = \boldsymbol{U}_m^H\boldsymbol{X}_m\boldsymbol{U}_m$ for (\ref{eq.prob_sdp}); this unitary rotation would not lead to performance loss in optimization. %In a nutshell, any feasible positive semidefinite matrix $\boldsymbol{X}_m$ for (\ref{eq.sdp0}) can be always decomposed into the structure of $\boldsymbol{U}_m\boldsymbol{\bar X}_m\boldsymbol{U}^H_m$.

%{\color{blue} Intuitively, this can be explained as follows. Assume that an $N_t\times N_t$ matrix $\boldsymbol{X}_m$ is feasible to (\ref{eq.sdp0}). Denote by $\boldsymbol{X}_m=\boldsymbol{U}\boldsymbol{\Sigma}\boldsymbol{U}^H$ its eigenvalue decomposition. If there are $N_t-1$ eigendirections in $\boldsymbol{U}$ are orthogonal to the channel $\boldsymbol{h}_m$, then $\boldsymbol{X}_m$ will meet the constraints (\ref{eq.ortho1}) naturally, and so is also feasible to (\ref{eq.prob_sdp}). However, such property may not always hold. Note that unitary rotation is invertible and does not change its trace value. Then there exists a feasible matrix to (\ref{eq.prob_sdp}) achieving the same performance as $\boldsymbol{X}_m$ to (\ref{eq.sdp0}) provided that $\boldsymbol{X}_m$ can be unitarily rotated to render $N_t-1$ eigendirections orthogonal to $\boldsymbol{h}_m$. Now the problem becomes whether all feasible matrices to (\ref{eq.sdp0}) can be rotated such that the resultant matrix has $N_t-1$ eigendirections orthogonal to $\boldsymbol{h}_m$. The answer is yes. The key here is that only $N_t-1$ eigendirections are required to be rotated to orthogonal to $\boldsymbol{h}_m$, and no constraint is imposed on the rest one direction. }

\end{remark}

Building on Lemma 1, we next propose an SDP based approach to obtaining the optimal REEL-BF solution.

\subsection{SDP based Solution to (\ref{eq.prob_qp})}

Let $\{\boldsymbol{X}^{\star}_m\}$ denote the optimal solution of (\ref{eq.sdp0}). By the equivalence between (\ref{eq.prob_sdp}) and (\ref{eq.sdp0}) in Lemma 1, we readily have the optimal solution for (\ref{eq.prob_sdp}) given by:
\begin{equation*}
    \boldsymbol{\bar X}^{\star}_m=\boldsymbol{U}_m^H\boldsymbol{X}_m^\star\boldsymbol{U}_m, ~~\forall m.
\end{equation*}
Let $\boldsymbol{\bar X}^{\star}_m=\begin{bmatrix}
  \eta^\star_m& \boldsymbol{\xi}^{\star H}_m\\
  \boldsymbol{\xi}^{\star}_m & \boldsymbol{\Gamma}^\star_m
  \end{bmatrix}$, $\forall m$.
Then, we can show the positivity of $\eta^\star_m$, $\forall m$, as follows.

\begin{lemma}
In the optimal solutions $\{\boldsymbol{\bar X}^{\star}_m\}$ for (\ref{eq.prob_sdp}), it holds that $\eta^\star_m >0$, $\forall m$.
\end{lemma}

\begin{IEEEproof}
Suppose that (\ref{eq.sdp0}) is solvable. With the positive SINR target, i.e., $\gamma_m>0$, it follows that
\begin{equation}
\left[\boldsymbol{U}_m^H\boldsymbol{X}^\star_m\boldsymbol{U}_m\right]_{1,1}=\boldsymbol{\bar h}_m^H\boldsymbol{X}^\star_m\boldsymbol{\bar h}_m>0,~~\forall m,
\end{equation}
since $\boldsymbol{\bar h}_m^H\boldsymbol{X}^\star_m\boldsymbol{\bar h}_m$ is actually the numerator of the achieved SINR with $\boldsymbol{X}^\star_m$ for user $m$ [cf. (\ref{eq.lem_SINR})]. By the fact $\eta^\star_m = [\boldsymbol{U}_m^H\boldsymbol{X}^\star_m\boldsymbol{U}_m]_{1,1}$, the lemma follows.
\end{IEEEproof}

Given that $\eta^\star_m >0$, $\forall m$, we can then perform the following matrix decomposition for $\boldsymbol{\bar X}^{\star}_m$ such that
\begin{equation} \label{eq.decom}
  \boldsymbol{\bar X}^{\star}_m =\boldsymbol{u}_{m}\boldsymbol{u}_{m}^H+ \sum_{k=1}^{R_m}\boldsymbol{q}_{m,k}\boldsymbol{q}_{m,k}^H, ~~\forall m,
\end{equation}
where
\begin{equation}
\boldsymbol{u}_m = \begin{bmatrix}
\sqrt{\eta^{\star}_m} \\ {\boldsymbol{\xi}^{\star}_m}/{\sqrt{\eta^\star_m}}
\end{bmatrix}, ~~
\begin{bmatrix}
\boldsymbol{q}_{m,1},...,\,\boldsymbol{q}_{m,R_m}
  \end{bmatrix}=
\begin{bmatrix}
\boldsymbol{0}_{1\times R_m} \\ \boldsymbol{\bar \Omega}_m
\end{bmatrix},
\end{equation}
with $\boldsymbol{\bar \Omega}_m\boldsymbol{\bar \Omega}^H_m=\boldsymbol{\Gamma}^{\star}_m-{\boldsymbol{\xi}^{\star}_m\boldsymbol{\xi}^{\star H}_m}/{\eta^{\star}_m}$, $R_m := \text{rank}(\boldsymbol{\bar \Omega}_m\boldsymbol{\bar \Omega}^H_m)$, and $\boldsymbol{\bar \Omega}_m\in \mathbb{C}^{(N_t-1)\times R_m}$, according to the analysis in Sec.~IV-A.

Suppose for now that $R_m \leq K-1$, $\forall m$. (How to select $K$ to ensure this condition will be discussed in the next subsection.) Then the omitted rank constraints (\ref{eq.prob_rank}) are naturally satisfied, and the optimal beamforming design for (\ref{eq.prob_qp}) can be retrieved. Specifically, the solutions to (\ref{eq.prob_qp}) are obtained as
\begin{equation}
\left\{\begin{split}
& \alpha_m^\star = \sqrt{\eta^\star_m} \\
& \boldsymbol{\beta}_m^\star = \boldsymbol{\xi}^\star_m/\sqrt{\eta_m^\star} \\
& \boldsymbol{\Omega}_m^\star = \left[\boldsymbol{\bar \Omega}_m ~~ \boldsymbol{0}_{(N_t-1)\times (K-1-R_m)}\right],
\end{split}\right.
\end{equation}
for $m=1,...,M$. As a result, the optimal beamforming matrices for the users are given by
\begin{equation}
\boldsymbol{W}_m^\star =\boldsymbol{U}_m\begin{bmatrix}
\alpha_m^\star & \boldsymbol{0}_{1\times (K-1)}\\
\boldsymbol{\beta}_m^\star & \boldsymbol{\Omega}_{m}^\star
\end{bmatrix},~~\forall m.
\end{equation}

\subsection{Optimality Guarantees by Proper Selection of $K$}

We have shown that the optimal beamforming matrices $\{\boldsymbol{W}^\star_m\}$ can be computed from the SDP solution $\{\boldsymbol{X}^{\star}_m\}$ to (\ref{eq.sdp0}) (or, $\{\boldsymbol{\bar X}^{\star}_m\}$ to (\ref{eq.prob_sdp})) under the conditions $R_m = \text{rank}\left(\boldsymbol{\Gamma}^{\star}_m-{\boldsymbol{\xi}^{\star}_m\boldsymbol{\xi}^{\star H}_m}/{\eta^{\star}_m}\right) \leq K-1$, $\forall m$.

As Fact 1 states, the number $K$ of the beamformers per user is actually a design parameter that can be flexibly chosen from any positive integers. %A naive way to guarantee the tightness of the SDP relaxation (\ref{eq.prob_sdp}) or (\ref{eq.sdp0}) is: Start from $K=1$ and increase the value of $K$ in our LiST-BF design if needed, until the corresponding solution ranks satisfy $r_m  \leq K-1$, $\forall m$.
%Instead of such a potentially computation-intensive way,
We next show that the value of optimality-guaranteed $K$ can be in fact predetermined from the given problem structure.

Recall that $\boldsymbol{\Gamma}^{\star}_m-{\boldsymbol{\xi}^{\star}_m\boldsymbol{\xi}^{\star H}_m}/{\eta^{\star}_m}$, $\forall m$, are of size $(N_t-1) \times (N_t-1)$; hence, $R_m \leq N_t-1$, $\forall m$. A trivial result can be then established as follows.

\begin{lemma}\label{lem.Nt}
By selecting a $K \geq N_t$, the optimal REEL-BF solution for (\ref{eq.prob_qp}) can be retrieved from the solution of its SDP relaxation (\ref{eq.prob_sdp}) or (\ref{eq.sdp0}).
\end{lemma}

\begin{IEEEproof}
For any $K\geq N_t$, we have $R_m \leq N_t-1 \leq K-1$, $\forall m$. The result readily follows.
\end{IEEEproof}

Lemma 3 states that selection of $K>N_t$ would not increase the degree of freedom in the proposed REEL-BF design; i.e., $K$ should be always selected to be no more than $N_t$.

Furthermore, we can also determine the value of $K$ to guarantee optimality in accordance with the number of additional quadratic shaping constraints in (\ref{eq.sdp0}) or (\ref{eq.prob_sdp}). In \cite{Huang10,Law15}, the relationship between the solution rank profile of (\ref{eq.sdp0}) and the total number of constraints was investigated based on the polynomial-time rank reduction techniques. Building on the results in \cite{Huang10,Law15}, we have two lemmas stated as follows.

\begin{lemma} \label{lem.rank1}
Applying the rank reduction procedure \cite[Algorithm 1]{Huang10} to construct an optimal solution $\{\boldsymbol{X}_m^{\star}\}$ for (\ref{eq.sdp0}), it always holds that
\begin{align}\label{eq.lem_rank1}
\text{rank}(\boldsymbol{X}^{\star}_m)\leq \sqrt{L+PM+1},~~\forall m.
\end{align}
\end{lemma}

\begin{IEEEproof}
Please see Appendix A.
\end{IEEEproof}

\begin{lemma}\label{lem.rank2}
Suppose that $P=2$, and the double-sided constraints in (\ref{eq.sdp0}) are inactive at optimality or they reduce to $\boldsymbol{{C}}_{pm}\bullet\boldsymbol{X}_m\unrhd_{pm} 0$, $\forall p$, $m$, where $\unrhd_{pm}\in\{\geq,=\}$. Then the rank-one decomposition in \cite[Lemma 4.1]{Huang10} can be applied in the rank reduction procedure to construct a solution $\{\boldsymbol{X}_m^{\star}\}$ for (\ref{eq.sdp0}) such that
\begin{equation}\label{eq.lem_rank2}
 \text{rank}(\boldsymbol{X}_m^{\star}) \leq L+1, ~~\forall m.
\end{equation}
\end{lemma}

\begin{IEEEproof}
Please see Appendix B.
\end{IEEEproof}

These two lemmas can be viewed as extensions of the results in \cite{Huang10,Huang10b,Law15}. We highlight that, according to Lemma~\ref{lem.rank2}, up to $K-1+2M$ additional shaping constraints can be accommodated to guarantee rank-$K$ solutions for (\ref{eq.sdp0}). This can be better than the claim in Table~I of \cite{Law15} under certain conditions. For example, when $K=2$, the maximum allowable number of additional shaping constraints is $7$ for any $M$ in \cite{Law15}, while herein we can increase the number to 9 and 11 for $M=4$ and $M=5$, respectively, and the more $M$, the larger the number.

Lemmas \ref{lem.rank1} and \ref{lem.rank2} indicate that, given the $L+PM$ additional shaping constraints, one can obtain upper bounds for the solution ranks of (\ref{eq.sdp0}). This provides another avenue for setting $K$ value in our proposed scheme to achieve the SDP bound promised by (\ref{eq.sdp0}).
Incorporating also the trivial bound in Lemma \ref{lem.Nt}, we can establish the following theorem.

\begin{theorem} The optimal REEL-BF solution achieving the SDP bound can be efficiently obtained when any one of the following three conditions is satisfied:
\begin{itemize}
\item[i)]  $K \geq N_t$;
\item[ii)] $K \geq \sqrt{L+PM+1}+2$;
\item[iii)] $K \geq \min\{L+3, \sqrt{L+PM+1}+2\}$ under the prerequisite conditions in Lemma 5.
\end{itemize}
\end{theorem}

\begin{IEEEproof}
Please see Appendix C.
\end{IEEEproof}

\begin{remark}
Theorem 1 means that we can pre-select the number $K$ of the beamformers per user in our REEL-BF design, independent of $\{\boldsymbol{h}_m,\boldsymbol{A}_{ij}, \boldsymbol{C}_{pm},\tau_i, \ell_{pm},\mu_{pm}, \unrhd_i\}$, i.e., the specific channels and shaping constraints. The optimality-guaranteed $K$ can be simply determined from the numbers of shaping constraints $\{L,PM\}$ and/or the number of the BS transmit antennas $N_t$. Note that the conditions on $K$ in Theorem 1 are actually sufficient conditions for the optimality of the proposed REEL-BF design. The optimality-guaranteed $K$ value can be much smaller than the ones specified in Theorem 1, as shown by the numerical results in Sec.~V. It may be possible to derive a tighter sufficient condition by further exploiting the structure of the given beamforming problem, i.e., the specific $\{\boldsymbol{h}_m, \boldsymbol{A}_{ij}, \boldsymbol{C}_{pm},\tau_i, \ell_{pm},\mu_{pm}, \unrhd_i\}$. This could be an interesting direction to pursue in the future work.
\end{remark}

\subsection{The Proposed Algorithm}

We are now ready to propose an efficient SDP based solver for our beamforming design problem (\ref{eq.prob_qp}), as summarized in Algorithm 1.

\begin{algorithm}
\caption{ Proposed algorithm to obtain the optimal REEL-BF solution achieving the SDP bound}
%\begin{algorithmic}[1]
\label{alg:Framwork}
%\scriptsize
%\State
{\bf Initialization:} Select $K = \min \{\sqrt{L+PM+1}+2, N_t\}$, or $K = \min\{L+3, \sqrt{L+PM+1}+2, N_t\}$ with the prerequisite conditions in Lemma \ref{lem.rank2}.

%\State
{\bf Input:} The CSI $\boldsymbol{\bar h}_m=\boldsymbol{h}_m/\|\boldsymbol{h}_m\|$, $\boldsymbol{F}_m=\text{null}(\boldsymbol{\bar h}_m\boldsymbol{\bar h}_m^H)$, unitary matrix $\boldsymbol{U}_m=[\boldsymbol{\bar h}_m ~\boldsymbol{F}_m]$, $\forall m$.

%\State
\vspace{0.2cm}
{\bf Step 1:} Solve problem (\ref{eq.sdp0}) to obtain $\{\boldsymbol{X}_m^\star\}$, and let
\[ \boldsymbol{\bar X}_m^\star:=\begin{bmatrix}
  \eta^\star_m& \boldsymbol{\xi}^{\star H}_m\\
  \boldsymbol{\xi}^{\star}_m & \boldsymbol{\Gamma}^\star_m
  \end{bmatrix} =\boldsymbol{U}_m^H\boldsymbol{X}_m^\star\boldsymbol{U}_m,~~\forall m.\]

{\bf Step 2:}
Perform matrix decomposition for $\boldsymbol{\bar X}^\star_m$ such that
\[
  \boldsymbol{\bar X}^{\star}_m =\boldsymbol{u}_{m}\boldsymbol{u}_{m}^H+ \sum_{k=1}^{R_m}\boldsymbol{q}_{m,k}\boldsymbol{q}_{m,k}^H,~~\forall m,
\]
where $\boldsymbol{u}_m = \begin{bmatrix}
\sqrt{\eta^{\star}_m} \\ {\boldsymbol{\xi}^{\star}_m}/{\sqrt{\eta^\star_m}}
\end{bmatrix}$ and $[\boldsymbol{q}_{m,1},...,~\boldsymbol{q}_{m,R_m}]=
\begin{bmatrix}
\boldsymbol{0}_{1\times R_m} \\ \boldsymbol{\bar \Omega}_m\end{bmatrix}$ with $\boldsymbol{\bar \Omega}_m\boldsymbol{\bar \Omega}_m^H=\boldsymbol{\Gamma}^{\star}_m-{\boldsymbol{\xi}^{\star}_m\boldsymbol{\xi}^{\star H}_m}/{\eta^{\star}_m}$, $R_m=\text{rank}\left(\boldsymbol{\bar \Omega}_m\boldsymbol{\bar \Omega}_m^H\right)$, and $\boldsymbol{\bar \Omega}_m \in\mathbb{C}^{(N_t-1)\times R_m}$, $\forall m$.

%\State
{\bf Step 3:} Obtain the solution for (\ref{eq.prob_qp}): $\forall m$,
\begin{equation}\nonumber
\left\{\begin{split}
& \alpha_m^\star = \sqrt{\eta^\star_m} \\
& \boldsymbol{\beta}_m^\star = \boldsymbol{\xi}^\star_m/\sqrt{\eta_m^\star} \\
& \boldsymbol{\Omega}_m^\star = \left[\boldsymbol{\bar \Omega}_m~~ \boldsymbol{0}_{(N_t-1)\times (K-1-R_m)}\right].
\end{split}\right.
\end{equation}

%\State
{\bf Output:}
The optimal REEL-BF solution:
\begin{equation}\nonumber
\boldsymbol{W}_m^\star =\boldsymbol{U}_m\begin{bmatrix}
\alpha_m^\star & \boldsymbol{0}_{1\times (K-1)}\\
\boldsymbol{\beta}_m^\star & \boldsymbol{\Omega}_{m}^\star
\end{bmatrix},~~\forall m.
\end{equation}
%\end{algorithmic}
\end{algorithm}

By Theorem 1, Algorithm 1 is guaranteed to output the linear beamforming solution that achieves the SDP bound promised by the solution of (\ref{eq.sdp0}). This algorithm only involves solving an SDP, as well as performing unitary rotations and eigenvalue decompositions. Hence, the total computation cost is dominated by the interior-point SDP solver, with a worst-case computational complexity ${\cal O}(\sqrt{MN_t+L+2PM}M^{3}N_t^{6})$\cite{Wang14}.

Note that the complexity of Algorithm 1 in fact does not increase with the value of $K$. By simply choosing $K=N_t$, it is guaranteed to obtain the optimal REEL-BF solution that achieves the SDP bound. However, the BSs in the future cellular networks can be equipped with a very large number of antennas (e.g., massive MIMO in 5G systems); hence, setting $K=N_t$ could be very inconvenient for practical implementation of the resultant REEL-BF scheme, as the BS may need to generate too many redundant signals and perform too many (more than necessary) shaping beamforming. This motivates us to investigate other sufficient conditions through Lemmas 4 and 5 in order to determine the minimum possible $K$ value to guarantee the optimality of our REEL-BF scheme.
For implementation convenience, we select such a minimum $K$ value in the initialization step of Algorithm 1. %We note that, Theorem 1 accommodates the worst possible instance. In practice, the order $K$ value is usually much smaller

\begin{remark}
Overall, some comments are in order:
\begin{itemize}
\item The REEL-BF can deliver a low-complexity rank-$K$ beamforming scheme with full-rate transmission and without (block) encoding/decoding delays.

\item The SDP relaxation for the proposed REEL-BF scheme is equivalent to the one for general downlink beamforming problem.

\item The proposed REEL-BF scheme can always achieve the performance bound promised by its SDP relaxation.

\item Based on the SDP solver, the optimal REEL-BF solution can be efficiently obtained by Algorithm 1.
\end{itemize}
\end{remark}

\begin{remark}
With low-complexity implementations, the proposed REEL-BF scheme is a linear beamforming scheme. Theorem 1 establishes that the optimal REEL-BF solution is guaranteed to achieve the SDP bound in a simple and structured manner. It has been well known that the solution of SDP relaxation (\ref{eq.sdp0}) provides a lower bound for the transmission power with linear downlink beamforming schemes. It remains a gap between the performance of state-of-the-art beamforming schemes and such a lower bound in many scenarios. The proposed REEL-BF scheme closes this gap, providing an optimal solution for MU-MISO downlink beamforming design with arbitrary shaping constraints. %As SDP techniques have been widely employed in beamforming designs in other system models and setups, the proposed approach has a far-reaching implications.
\end{remark}

%%%%%%%%%%%%%%%%%%%%%%%%%%%%%%%%%%%%%%%%%%%%%%%%%%%%%%%%%%%%%%%%%%%%%%
%                                                                    %
%               Section: Numerical Results                                %
%                                                                    %
%%%%%%%%%%%%%%%%%%%%%%%%%%%%%%%%%%%%%%%%%%%%%%%%%%%%%%%%%%%%%%%%%%%%%%

\section{Numerical Results}
In this section, numerical results are provided to illustrate the REEL-BF performance for beamforming designs with additional context-specific shaping constraints. Consider a MU-MISO downlink, where the BS equipped with $N_t$ antennas serves $M$ single-antenna users. We set the same SINR target for all users, i.e., $\gamma_m=\gamma$, $\forall m$. The noise variance at these $M$ users is set to $\sigma^2_m=\sigma^2$, $\forall m$.

In all simulations, we select: i) the conventional rank-one beamforming scheme (labeled as ``Rank-one''), ii) the Alamouti-code based scheme (labeled as ``Alamouti-based''), and iii) the rate-3/4 order-4 OSTBC based scheme (``Rate-3/4-OSTBC''), as the baseline schemes for performance comparison. %The performance with the SDP (\ref{eq.sdp0}), labeled as ``SDP relaxation'', is also provided as a benchmark.
Note that the order-4 OSTBC has a rate of 3/4. For a fair comparison, its SINR target should be adjusted such that the same information rate is achieved. Suppose for simplicity that the achievable rate is given by Shannon's well-known formula: $r=\log_2(1+\gamma)$ (bits/s/Hz). Then, for the same information rate $r$, while the full-rate schemes require an SINR target $\gamma=2^r-1$, the Rate-3/4-OSTBC based scheme requires $\gamma=2^{\frac{4}{3}r}-1$. The CVX optimization package\cite{cvx} is used to solve the SDPs. We declare that $\text{rank}(\boldsymbol{X}^\star_m) = \kappa$ if the $(\kappa+1)$-th largest eigenvalue is smaller than $0.01\%$ of the sum of all eigenvalues.

\subsection{Design with EH Constraints}

In this subsection, we consider the beamforming design with external wireless charging terminals. We assume an EH scenario, where $M$ users and $L$ EH terminals are served by the BS. For simplicity, we assume the same EH target for the $L$ charging terminals, i.e., $\tau_l=\tau$ for $l=1,...,L$, and the noise variance $\sigma^2=0.1$.

Consider first the line-of-sight (LoS) transmission scenario, where $M=1$ (information decoding) user is located at direction $\theta_1=0^\circ$ relative to the BS array broadside. The minimal rate requirement is $r=1$ bit/s/Hz (corresponding to the SINR target $\gamma=0$ dB). A total of $110$ potential EH terminals are located in the directions
\begin{equation*}
[u_1,...,u_{110}]=[-90^\circ, -88.5^\circ,-87^\circ,...,-2^\circ,2^\circ,3.5^\circ,...,77^\circ],
\end{equation*}
relative to the serving BS under consideration. The LoS spatial signature is modeled as
\begin{equation} \label{eq.LoS_model}
\boldsymbol{h}(\theta) = \left[ 1, e^{j\pi\sin(\theta)},..., e^{j\pi(N_t-1)\sin(\theta)} \right]^T,
\end{equation}
where $\theta$ is the direction relative to the BS array broadside and the path loss of all users is assumed to be identical\cite{Huang10,Law15}.

\begin{figure}
  \centering
  %\vspace{-0.2cm}
  \includegraphics[width = 4.5in]{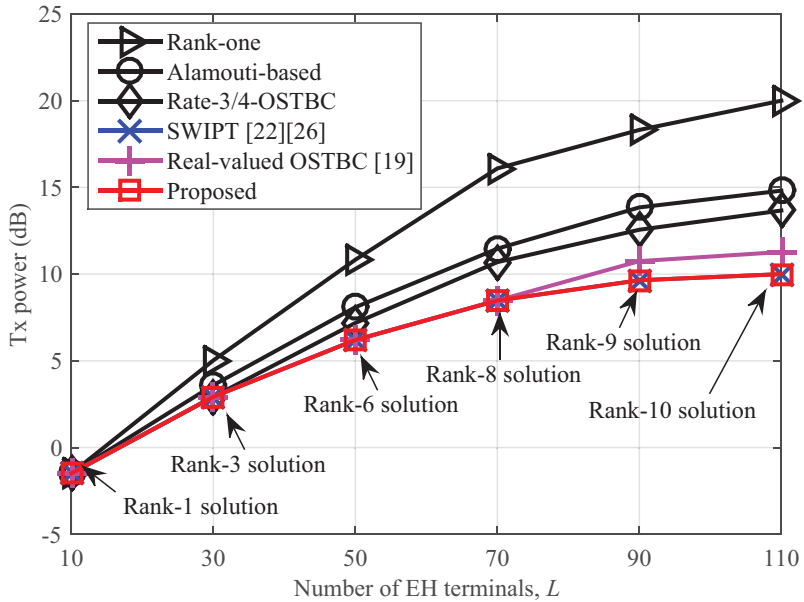}\\
    %\vspace{-0.3cm}
  \caption{Tx power vs. the number $L$ of EH terminals with $N_t=16$, $M=1$ and $\tau=10$ dB under LoS transmission scenario.} \label{fig.Power_vs_L_LoS}
    %\vspace{-0.3cm}
\end{figure}

Fig. \ref{fig.Power_vs_L_LoS} shows the transmit (Tx) power at the BS versus the number $L$ of EH terminals for different schemes with $N_t=16$ and $\tau=10$ dB, where the first $L$ EH terminals with directions $u_1,...,u_L$ are considered when $L$ value varies. In addition to the baseline schemes i)--iii), the performance of the real-valued OSTBC based scheme in \cite{Law15} and that of the SWIPT scheme based on the approach in \cite{Xu14,Li16} are also included. All the aforementioned schemes are developed based on SDP relaxation techniques. When the relevant SDP relaxation is not tight, a Gaussian randomization procedure is called to find an approximate solution, where the number of randomization instances is set to 100. It is verified by simulations that the proposed REEL-BF scheme always achieves the SDP bound (in this and all the remaining setups). Hence, the REEL-BF performance can serve as the optimal benchmark for all other schemes. As shown in Fig. \ref{fig.Power_vs_L_LoS}, the SWIPT scheme in \cite{Xu14,Li16} achieves the same performance with the REEL-BF scheme. This is expected since a similar transmission strategy is employed in \cite{Xu14,Li16}, although a different optimization approach is adopted. Clearly, the solution rank of (\ref{eq.sdp0}) grows with increasing $L$. The optimality cannot be always guaranteed for the real-valued OSTBC based scheme in \cite{Law15} for large $L$. For example, when $L=90$, the scheme in \cite{Law15} is suboptimal since the SDP relaxation (\ref{eq.sdp0}) yields a rank-9 solution (after rank reduction procedure); in this case, the proposed REEL-BF scheme achieves about 0.9 dB power gain over the real-valued OSTBC one. Observe that when $L=30$, even though the SDP solution rank is three, the Rate-3/4-OSTBC based scheme cannot achieve the SDP bound. This is because a higher SINR target needs to be adopted in its design due to the rate loss with the rate-3/4-OSTBC. As $L$ increases, the proposed REEL-BF scheme can have a significant performance gain over the baseline schemes (e.g., about 9.4 dB over Rank-one, 3.5 dB over Alamouti-based, and 2.3 dB over Rate-3/4-OSTBC based schemes at $L=70$).

\begin{figure}
  \centering
  \includegraphics[width = 4.5in]{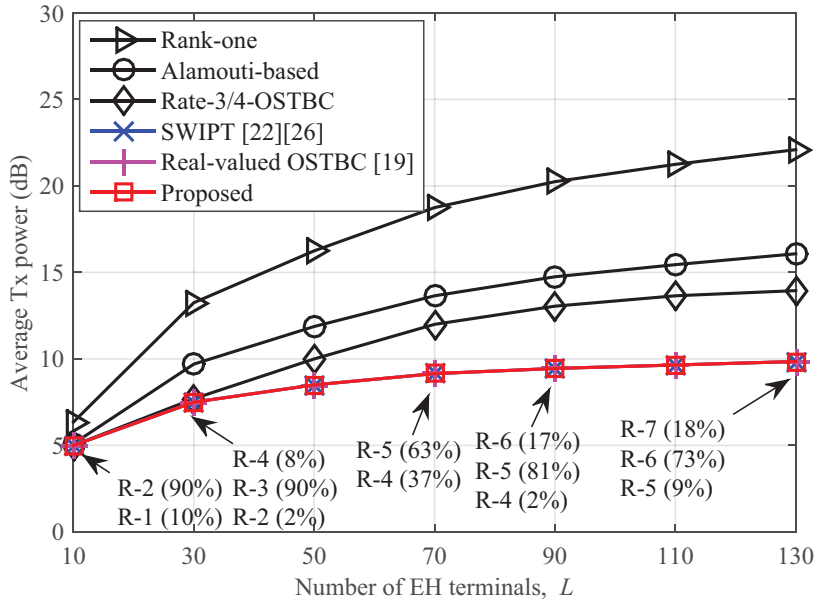}\\
  \caption{Average Tx power vs. number of EH terminals $L$ with $N_t=16$, $M=1$ and $\tau=10$ dB for Rayleigh fading scenario.} \label{fig.Power_vs_L_Rayleigh}
\end{figure}

We next consider Rayleigh fading downlink channels for $M=1$ information decoding user and $L$ EH terminals, i.e., $\boldsymbol{h}_i\sim {\cal CN}(\boldsymbol{0},\boldsymbol{I}_{N_t})$ for $i=1,...,M+L$, where the same path-loss and shadowing effects are assumed. The number of Monte-Carlo runs is set to 100 for randomized Rayleigh fading channel generation, and the number of Gaussian randomization instances is set to 100 for the baseline schemes if necessary. Fig.~\ref{fig.Power_vs_L_Rayleigh} shows the Tx power at the BS versus $L$ for different schemes. Again, as the number of EH terminals increases, the solution rank of (\ref{eq.sdp0}) increases, and the proposed REEL-BF scheme achieves significant performance gain over the baseline schemes. For example, when $L=30$, the solution rank is not larger than four, while the solution rank is greater than four with a probability of $98\%$ when $L=90$. Correspondingly, the gains of the proposed REEL-BF scheme over the Rank-one, Alamouti-based, and Rate-3/4-OSTBC based schemes are about 6.2 dB, 2.5 dB, and 0.1 dB for $L=30$, and about 11.8 dB, 5.6 dB, and 3.7 dB for $L=90$, respectively. Notice that the solution rank of the SDP (\ref{eq.sdp0}) is not greater than eight (after rank reduction procedure) for all $L\in[10,110]$. As a result, the real-valued OSTBC based scheme in \cite{Law15} always achieves the SDP bound as with the REEL-BF scheme. As with the LoS scenario, it can be seen in Fig.~\ref{fig.Power_vs_L_Rayleigh} that the SWIPT scheme \ref{Xu14,Li16} again achieves the optimal performance, i.e., the SDP bound.
%Different from the LoS scenario, the SWIPT scheme performs inferiorly to the REEL-BF scheme.
%This is because the resultant SDP relaxation from direct application of the approach in \cite{Xu14} is in general not tight for power minimization with EH constraints in this Rayleigh fading setup; in many cases a Gaussian randomization procedure is called to obtain a suboptimal solution, leading to significant performance loss.

%\begin{table}[t]
%\small
%\caption{Percentage distribution of the energy beamformer number for the SWIPT scheme \cite{Xu14} for different $L$.} \label{tab.num_energy_BF}
%\begin{center}
%\begin{tabular}{|c|c|c|c|c|c|c|}
%\hline  \multirow{2}{*} {number} & \multicolumn{6}{c|}{ $L$ }  \\
%\cline{2-7} & 10   & 30   &50    & 70   &110    & 130    \\
%\hline 1    & 34\% & 7\%  & 3\%  & 0    & 0     & 0\%    \\
%\hline 2    & 64\% & 25\% & 25\% & 19\% & 7\%   & 7\%   \\
%\hline 3    & 1\%  & 65\% & 31\% & 33\% & 38\%  & 26\%   \\
%\hline 4    & 1\%  & 3\%  & 40\% & 32\% & 22\%  & 32\%   \\
%\hline 5    & 0    & 0    & 1\%  & 16\% & 22\%  & 19\% \\
%\hline 6    & 0    & 0    & 0    &  0   & 11\%  & 15\% \\
% \hline
%\end{tabular}
%\end{center}
%\end{table}

\subsection{Design with General Co-channel Interference Constraints}

Assume that the BS is equipped with a uniform linear array of $N_t=18$ antennas, and the antennas are spaced half wavelength apart. Following the parameter setting in the example 2 in \cite{Law15}, we consider $M=3$ users served by the BS and $L=19$ co-channel users connected to the neighboring BS(s). These $M=3$ user are located at $[\theta_1, \theta_2, \theta_3]= [-5^\circ, 10^\circ, 25^\circ]$, while the $L=19$ co-channel users are located at
\begin{subequations}
\begin{align*}
[\theta_4,\theta_5,...,\theta_{22}] = [ -89.375^\circ, -80^\circ, -70.625^\circ, -61.25^\circ,\\
 -51.875^\circ, -42.5^\circ, -33.125^\circ, -23.75^\circ, -14.375^\circ, 2^\circ, 3^\circ,\\
 17^\circ, 18^\circ, 34.375^\circ,  43.75^\circ, 53.125^\circ, 62.5^\circ, 71.875^\circ, 81.25^\circ
]
\end{align*}
\end{subequations}
relative to the BS array broadside. The LoS channel model is the same as defined in (\ref{eq.LoS_model}).
%For all users, the line-of-sight spatial signatures are modeled as
%\begin{equation} \label{eq.LoS_model}
%\boldsymbol{h}(\theta_i) = \left[ 1, e^{j\pi\sin(\theta_i)},..., e^{j\pi(N_t-1)\sin(\theta_i)} \right]',~~\forall i,
%\end{equation}
%where the path loss of all user is assumed to be identical \cite{Huang10}.

The interference power at direction $\theta_j$ is given by
\begin{equation}
f(\theta_j) = \sum_{m=1}^M \boldsymbol{A}_{jm}\bullet\boldsymbol{X}_m,~~j=4,...,22,
\end{equation}
where $\boldsymbol{A}_{jm}=\boldsymbol{h}(\theta_j)\boldsymbol{h}^H(\theta_j)$, $\forall m$. Assume that the interference power needs to limited by $\tau_j=0.1$, i.e., $f(\theta_j) \leq 0.1$ for $j=4,...,22$, which are typical co-channel interference constraints in e.g., cognitive radio scenario \cite{Zhang10}. To guarantee that the interference power attains a local minimal value at the direction $\theta_j$, we further impose the interference derivative constraints (c.f., \cite{Law15}):
\begin{equation}\label{eq.derivative}
-\epsilon \leq \frac{d f(\theta_j)}{d\theta_j} \leq \epsilon,~ ~\frac{d^2 f(\theta_j)}{d\theta_j^2} >0, ~~j=4,...,22,
\end{equation}
where the threshold is set to $\epsilon=10^{-5}$. Note that the constraints (\ref{eq.derivative}) can be written as the quadratic shaping constraints as in (\ref{eq.soft}). These interference derivative constraints would entail a high-rank beamforming design to achieve the SDP bound.

\begin{figure}
  \centering
  \includegraphics[width = 4.5in]{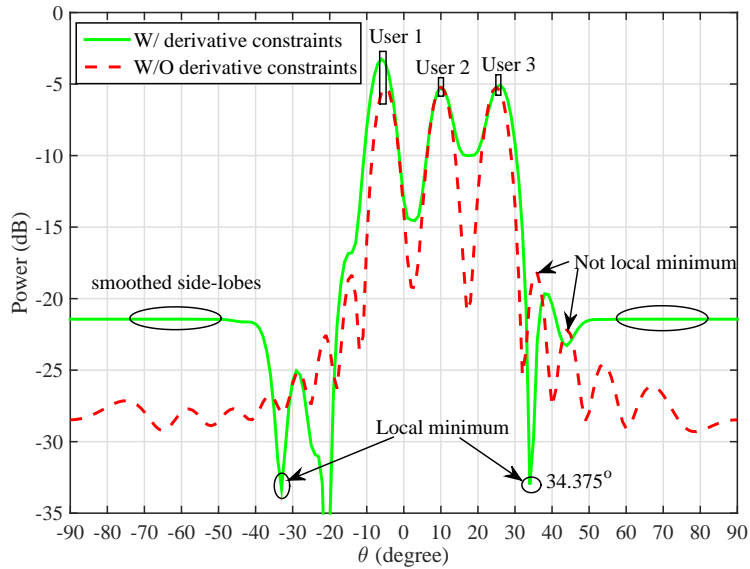}\\
  \caption{Sum BPs of the BS for different kinds of interference constraints. The number of the BS antennas is 18, and the user rate is $r=2$ bits/s/Hz. } \label{fig.BP_Int}
\end{figure}

%\begin{table}[htdp]
%\normalsize
%\caption{Rank Property for (\ref{eq.sdp0}) with $\gamma=5$ and $N_t=18$.}\label{tab.rank_Int}
%\vspace{-0.2cm}
%\begin{center}
%\begin{tabular}{|c|c|c|c|}
%\hline  & $\boldsymbol{X}^{\star}_{1}$ & $\boldsymbol{X}^{\star}_{2}$ & $\boldsymbol{X}^{\star}_{3}$\\
%\hline original rank of (\ref{eq.sdp0}) &  14 & 15 & 15 \\
% \hline reduced rank &  4 & 5 & 4\\
% \hline
%\end{tabular}
%\end{center}
%\vspace{-0.2cm}
%\end{table}

To see it, we evaluate the beam pattern (BP) of the BS, for $\theta\in [-90^\circ, ~90^\circ]$, according to
\begin{equation}
P(\theta )=\sum_{m=1}^M \boldsymbol{h}(\theta)\boldsymbol{h}^H(\theta) \bullet \boldsymbol{W}^\star_m\boldsymbol{W}^{\star H}_m,
\end{equation}
where $\{ \boldsymbol{W}_m^\star\} $ is provided by the REEL-BF solution. The BPs with and without interference derivative constraints are depicted in Fig.~\ref{fig.BP_Int}. Without the derivative constraints in (\ref{eq.derivative}), the sidelobes of the BP (displayed by dashed line) are only upper bounded and there is no specification on the shape of beam patterns. In this case, it can be proven that rank-one transmit covariance for each user is optimal \cite{Zhang10}. On the other hand, with (\ref{eq.derivative}), the requirement of locally minimum power at certain interference direction in fact imposes a shape requirement in the neighborhood of this interference direction. This usually leads to a more complicated beam pattern, as shown in Fig.~\ref{fig.BP_Int}. A high-rank transmit covariance matrix would be then required for each user; i.e., such a beam pattern cannot be realized by rank-one beamforming schemes due to lack of degrees of freedom in the design. On the other hand, with redundant-signal-aided shaping beamformers, more degrees of freedom are available in the beamforming design such that we could massage the co-channel interference levels within the allowable range to satisfy the desired local minimal requirements.

%{\color{blue} The rank profile for the optimal solution yielded by the CVX solver is: $\text{rank}(\boldsymbol{X}_1^\star)=14$, $\text{rank}(\boldsymbol{X}_2^\star)=15$, and $\text{rank}(\boldsymbol{X}_3^\star)=15$. After applying the polynomial-time rank reduction procedure \cite[Algorithm 1]{Huang10}, we can have: $\text{rank}(\boldsymbol{X}_1^\star)=2$, $\text{rank}(\boldsymbol{X}_2^\star)=3$, and $\text{rank}(\boldsymbol{X}_3^\star)=4$. It is clear that a high-rank transmit covariance matrix would be then required for each user to facilitate the desirable beam pattern with local minima ate certain interference direction; i.e., the redundant-signal-aided shaping beamformers are necessary to produce such a beam pattern.}

\begin{figure}
  \centering
  \includegraphics[width = 4.5in]{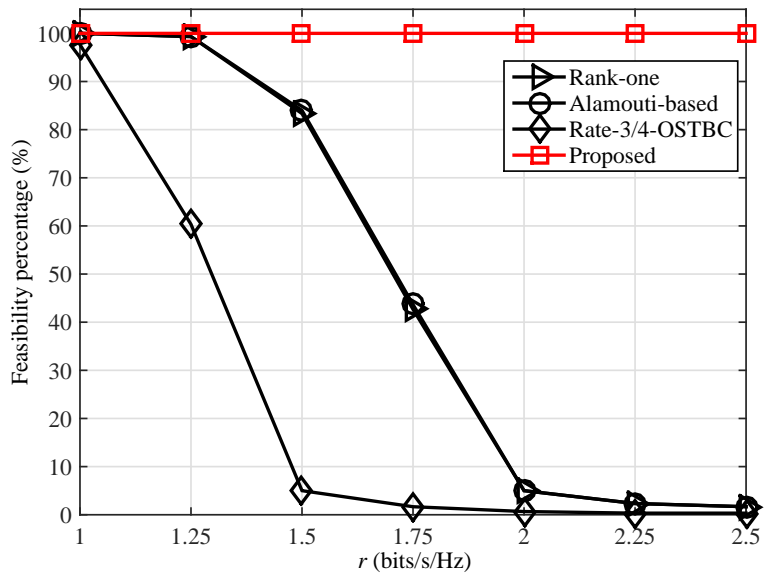}\\
  \caption{The feasibility percentage vs. rate of all schemes with general co-channel interference constraints under LoS transmission scenario.} \label{fig.Fea_Int}
\end{figure}

To see it, we now assume that the angles of departure at the BS for all the users are subject to variation in different Monte-Carlo runs; i.e.,
\begin{equation}
{\tilde \theta_i}= \theta_i + \Delta\theta_i,~~\forall i=1,...,22,
\end{equation}
where $\Delta\theta_i$ are drawn from a uniform distribution within the interval $[-0.25^\circ, ~0.25^\circ]$. The results are averaged over 300 independent Monte-Carlo runs. Note that the SWIPT scheme in \cite{Xu14} is not applicable to this scenario, whereas the real-valued OSTBC based scheme in \cite{Law15} always achieves the SDP bound as with the proposed REEL-BF scheme. Hence, we only use the baseline schemes i)--iii) for performance comparison in ensuing simulations. For the baseline schemes, the Gaussian randomization procedure might fail to yield a feasible low-rank solution to satisfy all the constraints. Fig.~\ref{fig.Fea_Int} depicts the feasibility percentage versus the transmission rate $r$. Observe that the proposed REEL-BF scheme is always feasible for the rate regime of $[1,~2.5]$~bits/s/Hz, while the feasibility percentage of the baseline schemes decreases with increasing rate requirement. The Rank-one scheme exhibits a similar feasibility trend as the Alamouti-based one. Specifically, when $r=1$~bit/s/Hz, both the Rank-one and Alamouti-based schemes are always feasible (and optimal), since the SDP relaxations always admit rank-one solutions. As $r$ increases, the rank of the optimal solution increases, and these two schemes fail to yield a feasible solution in many cases. Due to the higher SINR requirements caused by its rate loss, the Rate-3/4-OSTBC based scheme exhibits a significantly reduced feasibility percentage when compared to the other schemes. For example, when $r$ increases from $1.25$~bits/s/Hz to $1.5$~bits/s/Hz, the feasibility percentages for the Alamouti-based and the Rate-3/4-OSTBC based schemes decrease from about $99\%$ and $60\%$ to about $83\%$ and $6\%$, respectively. In a nutshell, the REEL-BF scheme with an appropriate $K$ value is always feasible and optimal, whereas the suboptimal baseline schemes would become infeasible for large $r$.

%\begin{table}[htdp]
%\small
%\caption{Percentage distribution of  $(\text{rank}(\boldsymbol{X}^\star_1),~\text{rank}(\boldsymbol{X}^\star_2),~\text{rank}(\boldsymbol{X}^\star_3))$ for (\ref{eq.sdp0}) with different $r$.} \label{tab.rank_vs_rate}
%\vspace{-0.35cm}
%\begin{center}
%\begin{tabular}{|c|c|c|c|c|}
%\hline  \multirow{2}{*} {rank} & \multicolumn{4}{c|}{ $r$ (bits/s/Hz) }  \\
%\cline{2-5} & 1 & 1.5 & 2 & 2.5 \\
%\hline 1 & (100, 100, 100) & (83, 83, 83) & (5, 5, 5) & (2, 2, 2) \\
%\hline 2 & (0, 0, 0) & (0, 1, 1) & (0, 0, 0) &(0, 0, 0)\\
%\hline 3 & (0, 0, 0) & (2, 1, 1) & (1, 0, 1)  & (0, 0, 1)\\
%\hline 4 & (0, 0, 0) & (8, 11, 12) & (68, 58, 65) & (67, 46, 60)\\
%\hline 5 & (0, 0, 0) & (7, 4, 3) & (26, 37, 29) & (31, 52, 37)\\
% \hline
%\end{tabular}
%\end{center}
%\vspace{-0.2cm}
%\end{table}

\subsection{Design with Additional Relaxed-Nulling Constraints}
Besides the EH and typical co-channel interference constraints, we next investigate the beamforming design with additional relaxed-nulling constraints in (\ref{eq.sdp0}).

In addition to $M=2$ users and $L$ EH terminals, we now also consider $F=10$ co-channel users served by the neighboring BS(s) and $P=2$ co-channel users which require relaxed nulling. The Rayleigh fading channels from the BS to these co-channel users are denoted as $\boldsymbol{f}_j\sim{\cal CN}(\boldsymbol{0},\boldsymbol{I}_{N_t})$, $\forall j=1,...,F$, and $\boldsymbol{g}_p\sim{\cal CN}(\boldsymbol{0},\boldsymbol{I}_{N_t})$, $\forall p=1,..,P$. The interference thresholds for $F$ co-channel users are set to 0.01, i.e.,
\begin{equation}
\sum_{m=1}^M \boldsymbol{f}_j\boldsymbol{f}^H_j \bullet \boldsymbol{W}_m\boldsymbol{W}_m^H \leq 0.01,~~\forall j=1,...,F.
\end{equation}
The relaxed nulling is to limit the interference caused by user $m$ to a fraction $\varsigma$ of the worst-case interference, i.e., $\|\boldsymbol{g}^H_p\boldsymbol{W}_m\|^2\leq \varsigma\|\boldsymbol{g}_p\|^2\|\boldsymbol{W}_m\|^2$ \cite{Ham06}, which can be written in a double-sided individual shaping form of (\ref{eq.double-sided}):
\begin{equation*}
0\leq \text{tr}(\boldsymbol{C}_{pm}\boldsymbol{W}_m\boldsymbol{W}_m)\leq +\infty, ~~\forall p,~m,
\end{equation*}
with $\boldsymbol{C}_{pm}=\varsigma\boldsymbol{I}_{N_t}-{\boldsymbol{g}_p\boldsymbol{g}_p^H}/{\|\boldsymbol{g}_{p}\|^2}$. The classical nulling corresponds to the special case with $\varsigma=0$. Here, we set $\varsigma=0.02$.

\begin{figure}
  \centering
  \includegraphics[width = 4.5in]{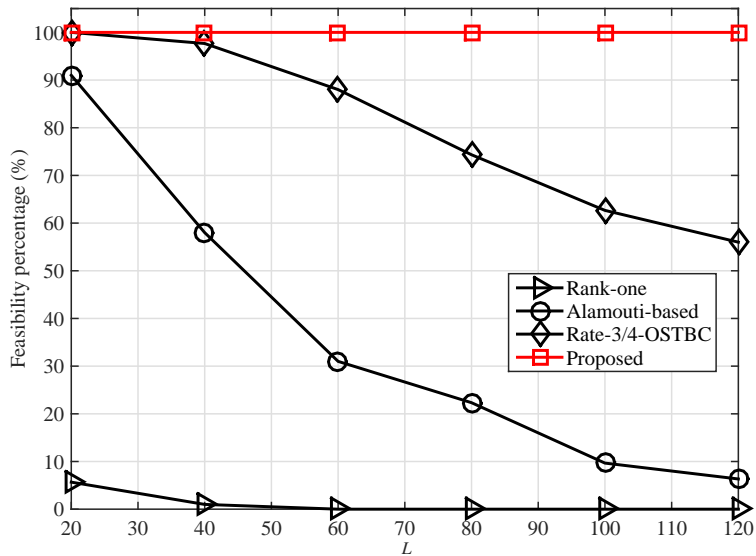}\\
  \caption{The feasibility percentage vs. number of EH terminals of all schemes with EH, typical co-channel interference, and relaxed-nulling constraints for Rayleigh fading scenario.} \label{fig.Fea_V_L_IntC}
\end{figure}

\begin{table} [t]%[htdp]
\small
\caption{Percentage distribution of $(\text{rank}(\boldsymbol{X}^\star_1),\text{rank}(\boldsymbol{X}^\star_2))$ vs. $L$.} \label{tab.rank_vs_L_IntC}
\begin{center}
\begin{tabular}{|c|c|c|c|c|}
\hline  \multirow{2}{*} {rank} & \multicolumn{4}{c|}{ $L$ }  \\
\cline{2-5} & 20 & 40 & 80 & 120 \\
\hline 1 & (25.7, 25) & (6, 8.3) & (0, 0) & (0, 0)\\
\hline 2 & (66.7, 66.7) & (54.3, 53) & (32.7, 32.3) &(13, 15)\\
\hline 3 & (7.6, 8.3) & (33, 31.7) & (33, 32)  & (23, 21)\\
\hline 4 & (0, 0) & (6.7, 7) & (34.3, 35.7) & (57.7, 58.3)\\
\hline 5 & (0, 0) & (0, 0) & (4, 2.3) & (6.3, 5.7)\\
 \hline
\end{tabular}
\end{center}
\end{table}

With addition of the relaxed-nulling constraints, again the Gaussian randomization procedure might fail to yield a feasible low-rank solution for the baseline schemes. Fig.~\ref{fig.Fea_V_L_IntC} demonstrates the feasibility percentage versus $L$ for all schemes with $r=2.06$~bits/s/Hz (corresponding to $\gamma=5$~dB), $\tau=10$~dB, $M=2$, and $N_t=16$, while Tab.~\ref{tab.rank_vs_L_IntC} gives the rank profile for the solutions to (\ref{eq.sdp0}). It is seen in Fig.~\ref{fig.Fea_V_L_IntC} that the proposed REEL-BF always yields feasible and optimal solutions for different $L$. By contrast, the feasibility percentage of the baseline schemes decreases with increasing $L$. When $L\in[20, 40]$, all schemes are feasible. For a large $L$ value (e.g., $L\geq 80$), the rank of the optimal solution becomes large with an increasing percentage and it is always larger than 1. As expected, the Rank-one scheme becomes infeasible and the feasibility percentage of other baseline schemes reduces.

Different from the problems with only EH constraints, the SDPs with additional $F+PM$ interference-control constraints usually admit low-rank solutions with a high probability. This confirms that Theorem 1 only provides sufficient conditions on the optimality-guaranteed $K$ value for the REEL-BF design. For the beamforming designs with interference-control type of shaping constraints, the SDP solution rank can be much smaller than the $K$ value specified by Theorem 1, i.e., a smaller number of the beamformers for each user can be used to achieve the SDP bound.

%%%%%%%%%%%%%%%%%%%%%%%%%%%%%%%%%%%%%%%%%%%%%%%%%%%%%%                                                                    %
%               Section: Conclusions                                %
%                                                                    %
%%%%%%%%%%%%%%%%%%%%%%%%%%%%%%%%%%%%%%%%%%%%%%%%%%%%%

\section{Conclusions}

In this paper, we proposed a REEL-BF scheme for downlink beamforming designs with arbitrary shaping constraints. It was established that the REEL-BF scheme is guaranteed to achieve the well-known SDP bound for the transmission power with linear beamforming schemes in a low-complexity and structured manner. An efficient algorithm was developed to obtain the optimal REEL-BF solution. Extensive numerical results demonstrated that the proposed REEL-BF scheme has significant performance gains over the existing alternatives.

Our work provides a novel approach to closing the performance gap between the practical linear beamforming schemes and the potential SDP bound. As SDP techniques have been widely employed in beamforming designs in many other system models and setups, the proposed approach has a far-reaching implication.

%%%%%%%%%%%%%%%%%%%%%%%%%%%%%%%%%%%%%%%%%%%%%%%%%%%%%%                                                                    %
%               Section: Appendix                                %
%                                                                    %
%%%%%%%%%%%%%%%%%%%%%%%%%%%%%%%%%%%%%%%%%%%%%%%%%%%%%

\section*{Appendix}

\subsection{Proof of Lemma \ref{lem.rank1}}
%\begin{IEEEproof}
Based on \cite[Lemma 3.1]{Huang10}, we can apply the rank reduction procedure \cite[Algorithm 1]{Huang10} to construct an optimal solution $\{\boldsymbol{X}^\star_m\}$ for (\ref{eq.sdp0}) such that
\begin{equation}\label{eq.lem_rank1.2}
\sum_{m=1}^M \text{rank}^2(\boldsymbol{X}_m^{{\star}})\leq M+L+PM.
\end{equation}

It is easy to show that $\boldsymbol{X}^\star_m$ cannot be zero matrix, since otherwise at least one of the SINR constraints in (\ref{eq.sdp0}) is violated.
Hence, we have $\text{rank}(\boldsymbol{X}^\star_m)\geq 1$, $\forall m$. It then follows from (\ref{eq.lem_rank1.2}) that:
\begin{equation}
\text{rank}^2(\boldsymbol{X}^\star_m) + (M-1) \leq  M+L+PM,~~\forall m.
\end{equation}
This then readily implies
\begin{equation}
\text{rank}(\boldsymbol{X}^\star_m) \leq \sqrt{L+PM+1},~~\forall m.
\end{equation}
%\end{IEEEproof}

\subsection{Proof of Lemma \ref{lem.rank2}}
%\begin{IEEEproof}
According to \cite[Lemma 4.2]{Huang10}, we can apply the rank-one decomposition in \cite[Lemma 4.1]{Huang10} in the rank reduction procedure to return a solution $\{\boldsymbol{X}_m^{\star}\}$ for (\ref{eq.sdp0}) such that
\begin{equation}\label{eq.lem_rank2.2}
\sum_{m=1}^M \text{rank}(\boldsymbol{X}_m^{\star}) \leq M+L.
\end{equation}
Following the similar lines in Lemma 4, we must have all $\boldsymbol{X}_m^{\star}$, $\forall m$, to be non-zero. It then follows from (\ref{eq.lem_rank2.2}) that
\begin{equation}\label{eq.lem_rank2.1}
 \text{rank}(\boldsymbol{X}_m^{{\star}})+(M-1) \leq M+L,~~\forall m.
\end{equation}
Thus, we readily conclude that
\begin{equation}
\text{rank}(\boldsymbol{X}^\star_m)\leq L+1,~~\forall m.
\end{equation}
%This completes the proof of Lemma 6.
%\end{IEEEproof}

\subsection{Proof of Theorem 1}
%\begin{IEEEproof}
To ensure that the optimal REEL-BF scheme achieves the SDP bound with (\ref{eq.sdp0}), it suffices to guarantee (\ref{eq.prob_rank}), i.e.,
\begin{equation}\label{eq.tight}
R_m \leq K-1, ~~\forall m.
\end{equation}
By Lemma 3, this naturally holds when $K\geq N_t$. This proves i) of Theorem 1.

To show ii), we observe that, for $m=1,...,M$,
\begin{subequations} \label{eq.rm_rank1}
\begin{align}
R_m&=~\text{rank}(\boldsymbol{\Gamma}^\star_m-{\boldsymbol{\xi}_m\boldsymbol{\xi}_m^H}/{\eta_m}) \leq~ \text{rank}(\boldsymbol{\Gamma}_m^\star) +1\\  %b
&\leq~\text{rank}(\boldsymbol{\bar X}_m^\star) +1 \leq~\text{rank}(\boldsymbol{X}_m^\star) +1\\ %d
&\leq~\sqrt{L+PM+1}+1, %e
\end{align}
\end{subequations}
where (\ref{eq.rm_rank1}b) holds due to the unitary similarity between $\boldsymbol{\bar X}_m^\star$ and $\boldsymbol{X}_m^\star$, and (\ref{eq.rm_rank1}c) follows from Lemma \ref{lem.rank1}.

By (\ref{eq.rm_rank1}), it follows that (\ref{eq.tight}) holds when
\begin{equation} \label{eq.t2}
\sqrt{L+PM+1}+1 \leq K-1.
\end{equation}
Hence, the choice of $K$ with
\begin{equation}
K \geq \sqrt{L+PM+1}+2
\end{equation}
guarantees that the optimality of (\ref{eq.sdp0}) can be achieved by the proposed REEL-BF solution; this proves ii).

Under the prerequisite conditions of Lemma \ref{lem.rank2}, we can follow the similar lines in (\ref{eq.rm_rank1}) to show
\begin{equation}\label{eq.rm_rank2}
R_m\leq L+2, ~~\forall m.
\end{equation}
Combining (\ref{eq.rm_rank2}) and the inequality chain of (\ref{eq.rm_rank1}) under the general conditions, we thus have
\begin{equation}
R_m\leq \min\{L+2,\sqrt{L+PM+1}+1\},~~\forall m.
\end{equation}
Again, it then follows that (\ref{eq.tight}) holds when
\begin{equation}
\min\{L+2, \sqrt{L+PM+1}+1\} \leq K-1.
\end{equation}
Hence, the selection of $K$ with
\begin{equation}
K\geq \min\{L+3, \sqrt{L+PM+1}+2\}
\end{equation}
guarantees that the optimality of (\ref{eq.sdp0}) can be achieved by the REEL-BF solution; the proof of iii) is completed.
%\end{IEEEproof}

\vspace{0.2cm} \noindent {\bf Acknowledgement:} The authors would like to thank an anonymous reviewer for helpful suggestion on use of the redundant signal embedded transmission structure to simplify our SDP bound achieving beamforming design.

\vspace{0.2cm}
\end{document}